\documentclass[reprint, superscriptaddress, prb]{revtex4-1}
\usepackage[colorlinks=true,citecolor=blue,linkcolor=magenta]{hyperref}
\usepackage{lineno}
\usepackage{amsmath}
\usepackage{amsfonts}
\usepackage{stmaryrd}
\usepackage{amssymb}
\usepackage{graphicx}
\usepackage{dcolumn}
\usepackage[british]{babel}
\usepackage{bm}

\begin{document}
\title{Low-loss interconnects for modular superconducting quantum processors}
\author{Jingjing Niu}
\affiliation{Shenzhen Institute for Quantum Science and Engineering, Southern University of Science and Technology, Shenzhen 518055, China}
\affiliation{International Quantum Academy, Shenzhen 518048, China}
\affiliation{Guangdong Provincial Key Laboratory of Quantum Science and Engineering, Southern University of Science and Technology, Shenzhen 518055, China}
\author{Libo Zhang}
\affiliation{Shenzhen Institute for Quantum Science and Engineering, Southern University of Science and Technology, Shenzhen 518055, China}
\affiliation{International Quantum Academy, Shenzhen 518048, China}
\affiliation{Guangdong Provincial Key Laboratory of Quantum Science and Engineering, Southern University of Science and Technology, Shenzhen 518055, China}
\author{Yang Liu}
\affiliation{Shenzhen Institute for Quantum Science and Engineering, Southern University of Science and Technology, Shenzhen 518055, China}
\affiliation{International Quantum Academy, Shenzhen 518048, China}
\affiliation{Guangdong Provincial Key Laboratory of Quantum Science and Engineering, Southern University of Science and Technology, Shenzhen 518055, China}
\author{Jiawei Qiu}
\affiliation{Shenzhen Institute for Quantum Science and Engineering, Southern University of Science and Technology, Shenzhen 518055, China}
\affiliation{International Quantum Academy, Shenzhen 518048, China}
\affiliation{Guangdong Provincial Key Laboratory of Quantum Science and Engineering, Southern University of Science and Technology, Shenzhen 518055, China}
\author{Wenhui Huang}
\affiliation{Shenzhen Institute for Quantum Science and Engineering, Southern University of Science and Technology, Shenzhen 518055, China}
\affiliation{International Quantum Academy, Shenzhen 518048, China}
\affiliation{Guangdong Provincial Key Laboratory of Quantum Science and Engineering, Southern University of Science and Technology, Shenzhen 518055, China}
\author{Jiaxiang Huang}
\affiliation{Shenzhen Institute for Quantum Science and Engineering, Southern University of Science and Technology, Shenzhen 518055, China}
\affiliation{International Quantum Academy, Shenzhen 518048, China}
\affiliation{Guangdong Provincial Key Laboratory of Quantum Science and Engineering, Southern University of Science and Technology, Shenzhen 518055, China}
\author{Hao Jia}
\affiliation{Shenzhen Institute for Quantum Science and Engineering, Southern University of Science and Technology, Shenzhen 518055, China}
\affiliation{International Quantum Academy, Shenzhen 518048, China}
\affiliation{Guangdong Provincial Key Laboratory of Quantum Science and Engineering, Southern University of Science and Technology, Shenzhen 518055, China}
\author{Jiawei Liu}
\affiliation{Shenzhen Institute for Quantum Science and Engineering, Southern University of Science and Technology, Shenzhen 518055, China}
\affiliation{International Quantum Academy, Shenzhen 518048, China}
\affiliation{Guangdong Provincial Key Laboratory of Quantum Science and Engineering, Southern University of Science and Technology, Shenzhen 518055, China}
\author{Ziyu Tao}
\affiliation{Shenzhen Institute for Quantum Science and Engineering, Southern University of Science and Technology, Shenzhen 518055, China}
\affiliation{International Quantum Academy, Shenzhen 518048, China}
\affiliation{Guangdong Provincial Key Laboratory of Quantum Science and Engineering, Southern University of Science and Technology, Shenzhen 518055, China}
\author{Weiwei Wei}
\affiliation{Shenzhen Institute for Quantum Science and Engineering, Southern University of Science and Technology, Shenzhen 518055, China}
\affiliation{International Quantum Academy, Shenzhen 518048, China}
\affiliation{Guangdong Provincial Key Laboratory of Quantum Science and Engineering, Southern University of Science and Technology, Shenzhen 518055, China}
\author{Yuxuan Zhou}
\affiliation{Shenzhen Institute for Quantum Science and Engineering, Southern University of Science and Technology, Shenzhen 518055, China}
\affiliation{International Quantum Academy, Shenzhen 518048, China}
\affiliation{Guangdong Provincial Key Laboratory of Quantum Science and Engineering, Southern University of Science and Technology, Shenzhen 518055, China}
\author{Wanjing Zou}
\affiliation{Shenzhen Institute for Quantum Science and Engineering, Southern University of Science and Technology, Shenzhen 518055, China}
\affiliation{International Quantum Academy, Shenzhen 518048, China}
\affiliation{Guangdong Provincial Key Laboratory of Quantum Science and Engineering, Southern University of Science and Technology, Shenzhen 518055, China}
\author{Yuanzhen Chen}
\affiliation{Shenzhen Institute for Quantum Science and Engineering, Southern University of Science and Technology, Shenzhen 518055, China}
\affiliation{International Quantum Academy, Shenzhen 518048, China}
\affiliation{Guangdong Provincial Key Laboratory of Quantum Science and Engineering, Southern University of Science and Technology, Shenzhen 518055, China}
\affiliation{Department of Physics, Southern University of Science and Technology, Shenzhen 518055, China}
\author{Xiaowei Deng}
\affiliation{Shenzhen Institute for Quantum Science and Engineering, Southern University of Science and Technology, Shenzhen 518055, China}
\affiliation{International Quantum Academy, Shenzhen 518048, China}
\affiliation{Guangdong Provincial Key Laboratory of Quantum Science and Engineering, Southern University of Science and Technology, Shenzhen 518055, China}
\author{Xiuhao Deng}
\affiliation{Shenzhen Institute for Quantum Science and Engineering, Southern University of Science and Technology, Shenzhen 518055, China}
\affiliation{International Quantum Academy, Shenzhen 518048, China}
\affiliation{Guangdong Provincial Key Laboratory of Quantum Science and Engineering, Southern University of Science and Technology, Shenzhen 518055, China}
\author{Changkang Hu}
\affiliation{Shenzhen Institute for Quantum Science and Engineering, Southern University of Science and Technology, Shenzhen 518055, China}
\affiliation{International Quantum Academy, Shenzhen 518048, China}
\affiliation{Guangdong Provincial Key Laboratory of Quantum Science and Engineering, Southern University of Science and Technology, Shenzhen 518055, China}
\author{Ling Hu}
\affiliation{Shenzhen Institute for Quantum Science and Engineering, Southern University of Science and Technology, Shenzhen 518055, China}
\affiliation{International Quantum Academy, Shenzhen 518048, China}
\affiliation{Guangdong Provincial Key Laboratory of Quantum Science and Engineering, Southern University of Science and Technology, Shenzhen 518055, China}
\author{Jian Li}
\affiliation{Shenzhen Institute for Quantum Science and Engineering, Southern University of Science and Technology, Shenzhen 518055, China}
\affiliation{International Quantum Academy, Shenzhen 518048, China}
\affiliation{Guangdong Provincial Key Laboratory of Quantum Science and Engineering, Southern University of Science and Technology, Shenzhen 518055, China}
\author{Dian Tan}
\affiliation{Shenzhen Institute for Quantum Science and Engineering, Southern University of Science and Technology, Shenzhen 518055, China}
\affiliation{International Quantum Academy, Shenzhen 518048, China}
\affiliation{Guangdong Provincial Key Laboratory of Quantum Science and Engineering, Southern University of Science and Technology, Shenzhen 518055, China}
\author{Yuan Xu}
\affiliation{Shenzhen Institute for Quantum Science and Engineering, Southern University of Science and Technology, Shenzhen 518055, China}
\affiliation{International Quantum Academy, Shenzhen 518048, China}
\affiliation{Guangdong Provincial Key Laboratory of Quantum Science and Engineering, Southern University of Science and Technology, Shenzhen 518055, China}
\author{Fei Yan}
\affiliation{Shenzhen Institute for Quantum Science and Engineering, Southern University of Science and Technology, Shenzhen 518055, China}
\affiliation{International Quantum Academy, Shenzhen 518048, China}
\affiliation{Guangdong Provincial Key Laboratory of Quantum Science and Engineering, Southern University of Science and Technology, Shenzhen 518055, China}
\author{Tongxing Yan}
\affiliation{Shenzhen Institute for Quantum Science and Engineering, Southern University of Science and Technology, Shenzhen 518055, China}
\affiliation{International Quantum Academy, Shenzhen 518048, China}
\affiliation{Guangdong Provincial Key Laboratory of Quantum Science and Engineering, Southern University of Science and Technology, Shenzhen 518055, China}
\author{Song Liu}
\email{lius3@sustech.edu.cn}
\affiliation{Shenzhen Institute for Quantum Science and Engineering, Southern University of Science and Technology, Shenzhen 518055, China}
\affiliation{International Quantum Academy, Shenzhen 518048, China}
\affiliation{Guangdong Provincial Key Laboratory of Quantum Science and Engineering, Southern University of Science and Technology, Shenzhen 518055, China}
\author{Youpeng Zhong}
\email{zhongyp@sustech.edu.cn}
\affiliation{Shenzhen Institute for Quantum Science and Engineering, Southern University of Science and Technology, Shenzhen 518055, China}
\affiliation{International Quantum Academy, Shenzhen 518048, China}
\affiliation{Guangdong Provincial Key Laboratory of Quantum Science and Engineering, Southern University of Science and Technology, Shenzhen 518055, China}
\author{Andrew N. Cleland}
\affiliation{Pritzker School of Molecular Engineering, University of Chicago, Chicago IL 60637, USA}
\affiliation{Center for Molecular Engineering and Material Science Division, Argonne National Laboratory, Argonne IL 60439, USA}
\author{Dapeng Yu}
\affiliation{Shenzhen Institute for Quantum Science and Engineering, Southern University of Science and Technology, Shenzhen 518055, China}
\affiliation{International Quantum Academy, Shenzhen 518048, China}
\affiliation{Guangdong Provincial Key Laboratory of Quantum Science and Engineering, Southern University of Science and Technology, Shenzhen 518055, China}
\affiliation{Department of Physics, Southern University of Science and Technology, Shenzhen 518055, China}

\maketitle

\textbf{
Scaling is now a key challenge in superconducting quantum computing. One solution is to build modular systems in which smaller-scale quantum modules are individually constructed and calibrated, and then assembled into a larger architecture. This, however, requires the development of suitable interconnects. Here, we report low-loss interconnects based on pure aluminium coaxial cables and on-chip impedance transformers featuring quality factors up to $8.1 \times 10^5$, which is comparable to the performance of our transmon qubits fabricated on single-crystal sapphire substrate. We use these interconnects to link five quantum modules with inter-module quantum state transfer and Bell state fidelities up to 99\%. To benchmark the overall performance of the processor, we create maximally-entangled, multi-qubit Greenberger-Horne-Zeilinger (GHZ) states. The generated inter-module four-qubit GHZ state exhibits 92.0\% fidelity. We also entangle up to 12 qubits in a GHZ state with $55.8 \pm 1.8\%$ fidelity, which is above the genuine multipartite entanglement threshold of 1/2. These results represent a viable modular approach for large-scale superconducting quantum processors.
}

Superconducting quantum circuits are a leading candidate for building scalable quantum processors~\cite{Kjaergaard2020}, whose size has increased significantly over the past decade, from a handful of qubits to over a hundred qubits~\cite{Arute2019,Wu2021,Ball2021}. With the near-term promise of processors with thousands of qubits~\cite{Ball2021}, driven by interest in intermediate-scale applications~\cite{Preskill2018}, come formidable engineering challenges, including available wafer size, device yield and crosstalk~\cite{Burkhart2021,Awschalom2021}, all constraining the scalability of monolithic quantum processor designs. This suggests the desirability of developing alternative modular approaches~\cite{Awschalom2021,Jiang2007,Kimble2008,Monroe2014,Chou2018,LaRacuente2022}, where smaller-scale quantum modules are individually constructed and calibrated, then assembled into a larger architecture using quantum coherent interconnects. Similar strategies have been pursued in other quantum systems such as atoms~\cite{Ritter2012,Dordevic2021}, ions~\cite{Monroe2014} and spin qubits~\cite{Pompili2021}. Indeed, it can be argued that modular quantum computing schemes is likely the only viable approach to scaling up to very large qubit numbers in the near term~\cite{Awschalom2021}. However, the requirement for high quality interconnects that can transfer vulnerable quantum states between separate processors remains an outstanding challenge~\cite{Awschalom2021,Kimble2008}.


\begin{figure*}[t]
\begin{center}
	\includegraphics[width=0.8\textwidth]{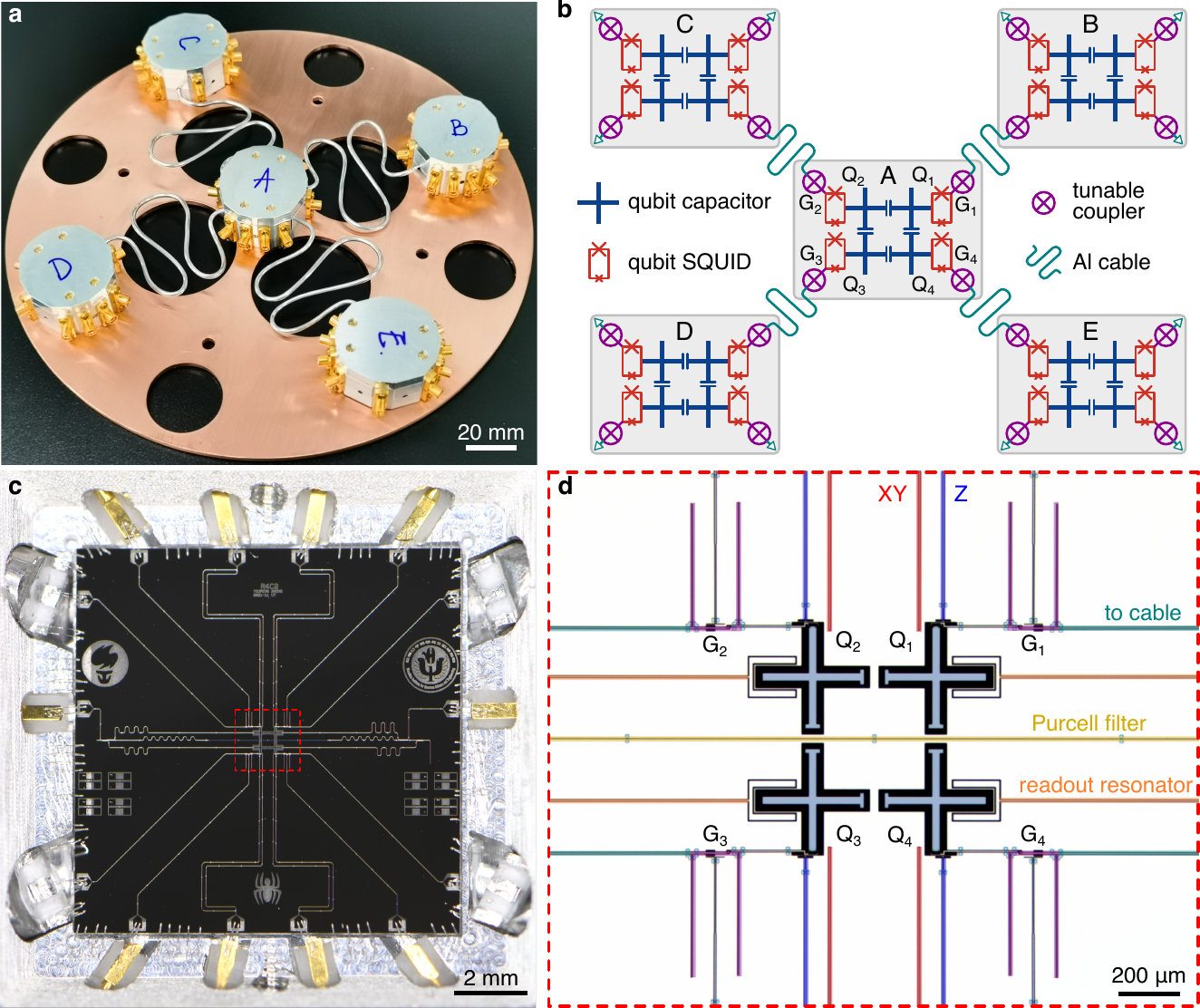}
	\caption{
    \label{fig1}
    {\bf Modular quantum processor design.}
    {\bf a,} Photograph of the modular quantum processor assembly.
    {\bf b,} Circuit schematic of the modular processor. Blue crosses represent the qubit capacitor pads; red rectangles with crosses represent the qubit junction loops (with asymmetric junction sizes); purple circles represent the tunable couplers, where the cross inside represents a Josephson junction; cyan serpentine lines represent the Al cables. The unused couplers are shorted to ground by bonding wires.
    {\bf c,} Photograph of the central module $A$, showing a spider-shaped circuit layout.
    {\bf d,} Detail micrograph of module $A$ in {\bf c}, showing the qubits, tunable couplers and control/readout circuitry.
    }
\end{center}
\end{figure*}
Reaching the fundamental loss limit of waveguides for transmitting signals has revolutionized our information society. For example, with the record-low loss of only 0.2 dB/km, optical fibers form the foundation of today's global telecommunication network.
The pursuit of low loss has not been unique to optical fibers, but also to radio frequency (RF) cables. With the rising of quantum information science, the quest for low-loss cables/waveguides has become imperative~\cite{Awschalom2021}.
Several recent experiments have demonstrated the connection of two superconducting quantum modules using niobium-titanium (NbTi) superconducting coaxial cables~\cite{Kurpiers2018,Axline2018,Campagne2018,Leung2019,Magnard2020,Zhong2021,Burkhart2021}.
Inter-module QST fidelities up to $91.1$\% have been reached, yet still not sufficient for large scale deployment.
Flip-chip schemes have also been investigated~\cite{Gold2021,Conner2021,Kosen2021}, achieving high fidelities while retaining many benefits of a modular architecture.
The routinely used NbTi cables have a linear loss of $\sim$5 dB/km (corresponding to a cable intrinsic quality factor of $Q_{cb}\sim 10^5$)~\cite{Kurpiers2018,Axline2018,Campagne2018,Zhong2021,Burkhart2021}.
In contrast to NbTi cables that have already been commercialized, pure aluminum (Al) cables have not yet been readily available, partially due to the low superconducting transition temperature $T_c$ of Al.
Here, we demonstrate coherent interconnects using low-loss coaxial cables made from pure Al, with Al wirebond connections to qubit processors.
While the $T_c=1.2$ K of Al is much lower than that of NbTi (9.7 K), surprisingly, the intrinsic quality factor $Q_{cb}$ of these Al cables at $\sim 10$~mK reaches $4.2\times 10^6$ -- equivalent to a record-low loss of 0.15 dB/km that is even lower than the loss value of 0.2 dB/km for optical fibers (see Supplementary Information for details).


\begin{figure*}[t]
\begin{center}
	\includegraphics[width=0.8\textwidth]{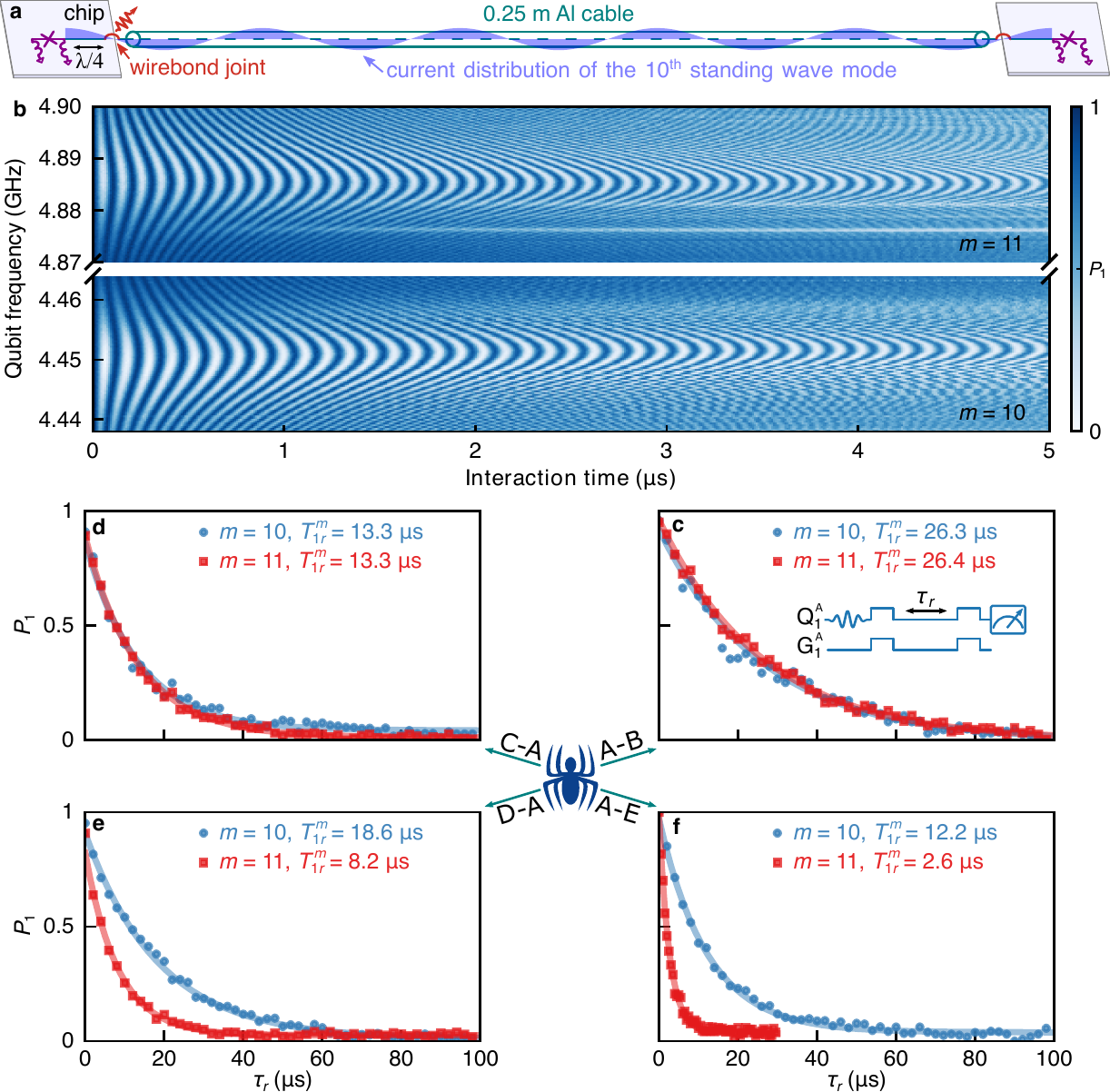}
	\caption{
    \label{fig2}
    {\bf Characterization of superconducting Al cable interconnects.}
    {\bf a,} Schematic of the chip-cable-chip interconnect (cyan), shorted to ground at both ends by the tunable couplers (purple). The current distribution of the 10$^{\rm th}$ standing wave mode is depicted (light blue). The on-chip CPW quarter-wavelength ($\lambda/4$) impedance transformer is designed to match the 10$^{\rm th}$ standing wave mode such that the wirebond joint (red) coincides with a current node, thus minimizing the current induced dissipation at the wirebond joint.
    {\bf b,} Vacuum Rabi oscillation between $Q_1^A$ and the $m=10$, 11 standing wave modes in the $A$-$B$ interconnect.
    {\bf c--f,} The lifetime $T_{1r}^{m}$ of the $m=10$, 11 standing wave modes in the four interconnects, measured using their associated qubits in module $A$. The 11$^{\rm th}$ mode in the $A$-$B$ interconnect has the highest internal quality factor, $Q_{\rm int}^m = \omega_m T_{1r}^{m}=8.1\times 10^5$. Inset to {\bf c}: Control pulse sequence for lifetime measurement. The spider symbol represents the central module $A$.
    }
\end{center}
\end{figure*}

Our modular quantum processor consists of five modules (see Fig.~\ref{fig1}), where each module $n$ ($n = A-E$) has four capacitively coupled transmon qubits $Q_i^n$ ($i=1$ to $4$), each qubit galvanically connected to a tunable coupler~\cite{Chen2014} $G_i^n$ for external connections. The five quantum modules are first individually calibrated at cryogenic temperatures to confirm that all components function properly. We then mount them on a circular copper plate and use four 0.25 m-long Al cables cut from a one-meter long cable to link the modules together in a star topology, with $A$ as the central module and $B$--$E$ on the periphery; see Fig.~\ref{fig1}{\bf a} and {\bf b}.  To reduce cable counts, each qubit's XY and Z control lines are combined using diplexers. To suppress dephasing noise while retaining some frequency tunability, we use asymmetric Josephson junctions with $\alpha=E_{J1}/E_{J2}=5.3$, where $E_{J1}$ and $E_{J2}$ are the Josephson energies of the two qubit junctions, giving a qubit frequency tuning range of $\sim 4.2$~GHz to $\sim 5.1$~GHz. More experimental details can be found in the Supplementary Information.

\begin{figure*}[t]
\begin{center}
	\includegraphics[width=0.8\textwidth]{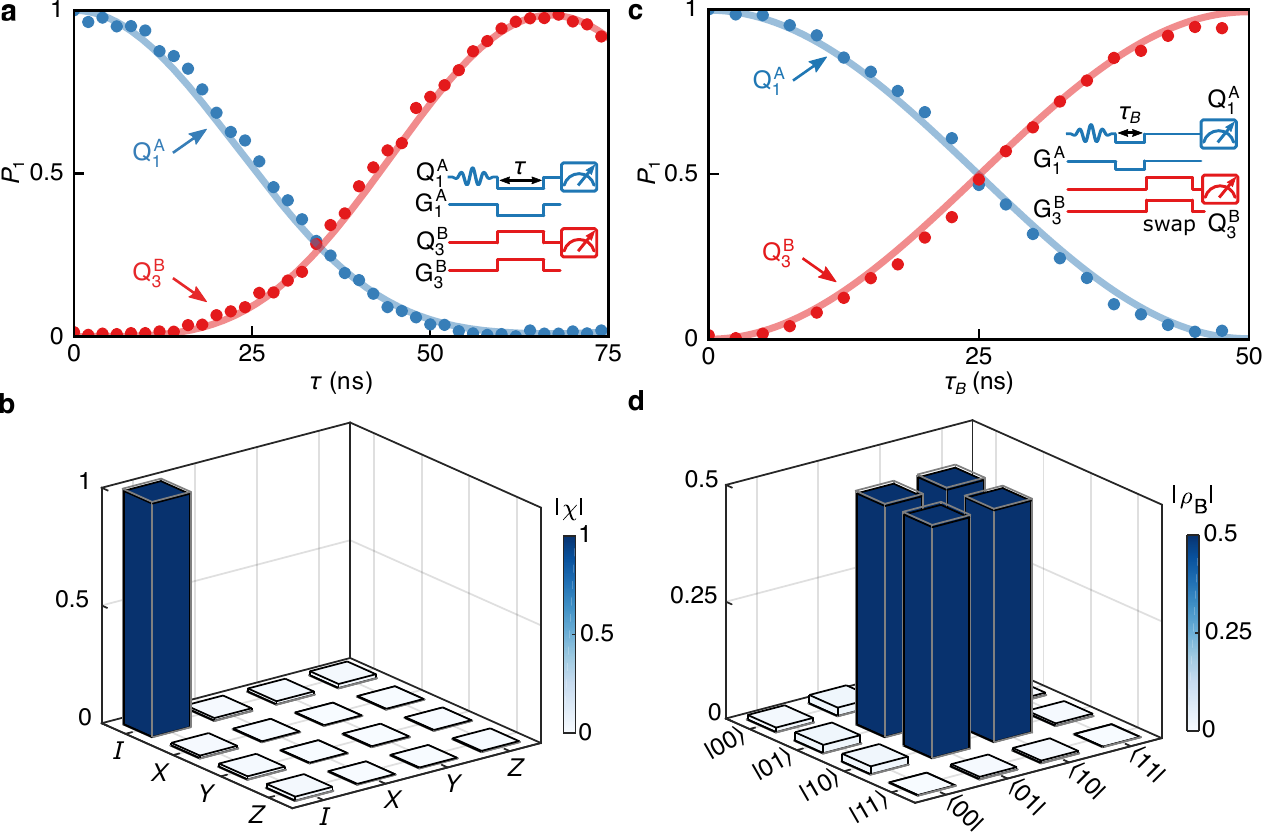}
	\caption{
    \label{fig3}
    {\bf High fidelity inter-module quantum state transfer and entanglement generation.}
    {\bf a,} Transferring a photon from $Q_1^A$ to $Q_3^B$.
    {\bf b,} Quantum process tomography of the state transfer shown in {\bf a} at $\tau=66$~ns, with a process fidelity of $\mathcal{F}_{QST}=99.1 \pm 0.5\%$. Horizontal axes are the Pauli operators $I$, $X$, $Y$ and $Z$.
    {\bf c,} Inter-module entanglement generation.
    {\bf d,} Density matrix of the Bell state generated in {\bf c} at $\tau_B=25$~ns, with a state fidelity of $\mathcal{F}_B =98.9\pm 0.6\%$.
    In {\bf a}, {\bf c}, the lines are simulations, the insets are control pulse sequences; in {\bf b}, {\bf d}, the solid bars and grey frames are the measured and ideal values respectively.
    }
\end{center}
\end{figure*}

We first tune up and characterize the interconnects. When the tunable couplers are tuned off, the coaxial cable is effectively shorted to ground by the tunable couplers at each end~\cite{Zhong2021}, supporting an evenly-spaced sequence of standing microwave modes, with a free spectral range $\omega_{\rm FSR}/2\pi \approx 0.44$~GHz determined by the cable length; see Fig.~\ref{fig2}{\bf a}. We now focus on the interconnect between modules $A$ and $B$. When we prepare the qubit $Q_1^A$ in its excited state $|1\rangle$ and subsequently turn on the coupler $G_1^A$, we observe vacuum Rabi oscillations between $Q_1^A$ and the cable standing wave modes when the qubit frequency matches with the $m$-th mode frequency $\omega_m/2\pi$~\cite{Zhong2021}; see Fig.~\ref{fig2}{\bf b}. The qubit frequency tuning range here allows access to the $m=10$ mode at 4.45~GHz and the $m=11$ mode at 4.885~GHz. The $\sim 6.5$~mm long on-chip CPW line is essentially a quarter-wavelength ($\lambda/4$) impedance transformer designed to match the 10$^{\rm th}$ standing wave mode, such that the wirebond joint coincides with a current node (as depicted in Fig.~\ref{fig2}{\bf a}), thus minimizing the current-induced dissipation at the wirebond joint; see Supplementary Information for details. We use $Q_1^A$ to measure the lifetime $T_{1r}^{m}$ of the two modes by first swapping a photon from $Q_1^A$ to the mode, waiting for a time $\tau_r$, then swapping out the photon. The population $P_1$ of the qubit excited state $|1\rangle$ is then measured, and decays exponentially with a time constant of $T_{1r}^{m}$; see Fig.~\ref{fig2}{\bf c}. Fitting the data gives $T_{1r}^{m}=26.3$~$\mu$s for the 10$^{\rm th}$ mode and $26.4$~$\mu$s for the 11$^{\rm th}$ mode respectively, comparable to the native coherence of the transmon qubits.
Other interconnects are characterized in the same way (see Fig.~\ref{fig2}{\bf d--f}), where the performance is somewhat non-uniform.
The best performance is found for the 11$^{\rm th}$ mode in the $A$-$B$ interconnect, with a standing mode internal quality factor $Q_{\rm int}^m=\omega_m T_{1r}^{m}=8.1\times 10^5$.
In comparison, state-of-the-art coplanar waveguides fabricated with ultra-high-vacuum molecular beam epitaxy (MBE) deposited Al on single-crystal sapphire substrate have achieved a quality factor of merely above $10^6$ at single photon levels~\cite{Megrant2012}.

Following Ref.~\cite{Zhong2021}, we use a hybrid scheme to transfer quantum states between module $A$ and $B$, using the 11$^{\rm th}$ standing wave mode $R$ and both the ``dark'' and ``bright'' eigenmodes of the $Q_1^A$-$R$-$Q_3^B$ tripartite system. This scheme balances the loss in the communication mode as well as in the qubits~\cite{Wang2012}.
We bias the couplers $G_1^A$ and $G_3^B$ to set the $Q_1^A$-$R$ coupling $g_1^A$ and the $Q_3^B$-$R$ coupling $g_3^B$ to the same strength $g_0/2\pi = 5$ MHz, meanwhile tuning both $Q_1^A$ and $Q_3^B$ to resonantly interact with $R$ for a duration $\tau$, see Fig.~\ref{fig3}{\bf a}. We note that while stronger coupling can be achieved here, extrinsic qubit loss could emerge as the coupling increases, exposing the qubits to loss channels at the wirebond joint~\cite{Zhong2021}. The coupling strength $g_0$ chosen here is similar to Ref.~\cite{Zhong2021}, which balances the intrinsic decoherence with the extrinsic loss. At $\tau=66$ ns, one photon is transferred from $Q_1^A$ to $Q_3^B$ with optimal fidelity. We perform quantum process tomography (see Supplementary Information) to characterize this QST process, yielding the process matrix $\chi$ shown in Fig.~\ref{fig3}{\bf b}, with a fidelity $\mathcal{F}_{QST} = \rm{Tr}(\chi \cdot \chi_{\mathcal{I}})=99.1\pm 0.5\%$, where $\chi_{\mathcal{I}}$ is the process matrix for the identity operation $\mathcal{I}$. Note all uncertainties reported here represent the standard deviation of repeated measurements. Numerical simulations give a process fidelity of $\mathcal{F}_{QST}=99.3\%$, agreeing well with the experiment.
Numerical analysis suggests this fidelity is mainly limited by the qubit decoherence; further improvement in QST fidelity requires a faster QST process and/or longer qubit coherence time.
In addition to transferring quantum states, we can also generate entanglement across modules, such as by preparing a Bell singlet state $|\psi_B\rangle=(|01\rangle+|10\rangle)/\sqrt{2}$ (here $|0\rangle$ is the qubit ground state) using a half-photon swap. We first turn on the $Q_1^A$-$R$ interaction for $\tau_B=25$~ns, sharing half a photon with $R$, then swap this entangled photon from $R$ to $Q_3^B$~\cite{Zhong2019}, a process we denote as QST/2 (see Fig.~\ref{fig3}{\bf c}). The density matrix $\rho_B$ of the resulting Bell state, obtained from quantum state tomography (see Supplementary Information), is shown in Fig.~\ref{fig3}{\bf d}, with a state fidelity of $\mathcal{F}_B = \langle \psi_B |\rho_B|\psi_B\rangle = 98.9\pm 0.6\%$, where numerical simulations give a very similar fidelity $\mathcal{F}_B=99.1\%$. The other interconnects give similar results, with an average QST fidelity of 98.2\% and Bell state fidelity of 98.6\% respectively; see Supplementary Information.

\begin{figure*}[t]
\begin{center}
	\includegraphics[width=0.8\textwidth]{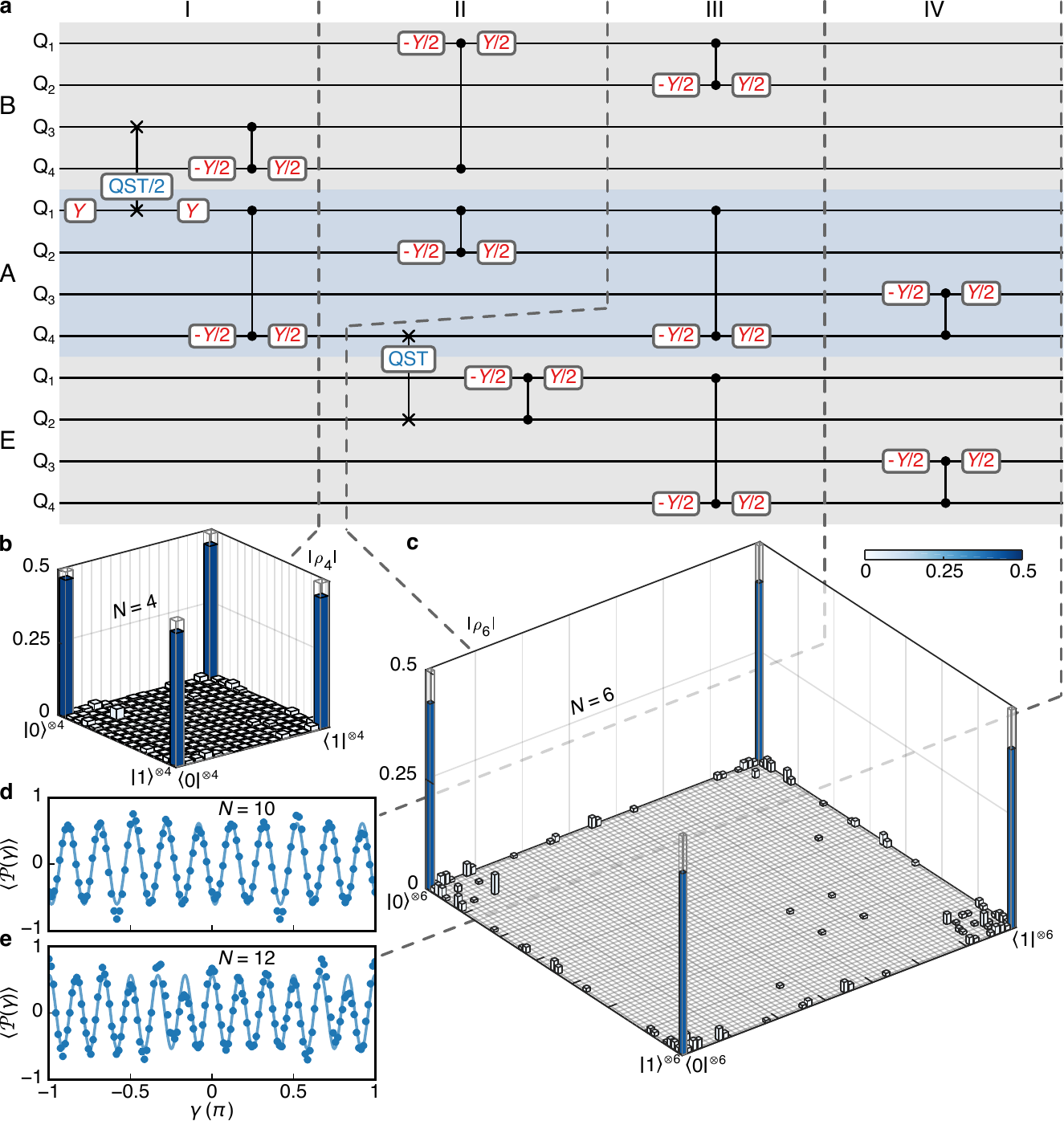}
	\caption{
    \label{fig4}
    {\bf Performance benchmarking using entanglement generation.}
    {\bf a,} The protocol for GHZ state generation. In step III, the QST from module $A$ to $E$ is moved forward to shorten the sequence length.
    {\bf b,} The 4-qubit GHZ state density matrix $\rho_4$ generated in step I, with fidelity $\mathcal{F}_{4} = 92.0\pm 0.6\%$.
    {\bf c,} The 6-qubit GHZ state density matrix $\rho_6$ generated in step II, with fidelity $\mathcal{F}_{6} = 83.2\pm 0.8\%$. For simplicity, density matrix elements smaller than 0.005 are not displayed.
    {\bf d, e,} The measured $\langle \mathcal{P}(\gamma) \rangle$ oscillation for the $N=10$ and $N=12$ GHZ states generated in step III and IV respectively. From the oscillation amplitude and the qubit joint-readout probabilities we estimate the GHZ state fidelity $\mathcal{F}_{10}=63.0\pm 1.5\%$ and $\mathcal{F}_{12}=55.8\pm 1.8\%$, respectively.
    The solid bars and grey frames in {\bf b}, {\bf c} are the measured and ideal values, respectively.
    }
\end{center}
\end{figure*}

With the qubits and interconnects carefully tuned up, we now benchmark the modular processor as a whole. Greenberger-Horne-Zeilinger (GHZ) states~\cite{Greenberger1990} are a fundamental resource in fault-tolerant quantum computation and quantum communication, and as they are highly susceptible to errors and decoherence, provide a sensitive benchmark for our processor. We use controlled-NOT (CNOT) gates to generate GHZ states step-by-step following the protocol in Fig.~\ref{fig4}{\bf a}, eventually entangling up to 12 qubits. The CNOT gates are composed of fast adiabatic controlled-Z (CZ) gates~\cite{Barends2014} and single qubit rotations, where the single qubit gates have an average fidelity of 99.85\% and the CZ gates have an average fidelity of 96.1\%; see Supplementary Information.
Starting from the Bell state in Fig.~\ref{fig3}{\bf d}, we apply CNOT gates on both sides simultaneously to first generate a 4-qubit GHZ state between modules $A$ and $B$, yielding $|\psi_{4}\rangle = (|0\rangle^{\otimes 4}+|1\rangle^{\otimes 4})\sqrt{2}$. The measured density matrix $\rho_4$ of this state is shown in Fig.~\ref{fig4}{\bf b}, with a fidelity $\mathcal{F}_{4} = \langle \psi_{4}|\rho_4|\psi_{4}\rangle = 92.0 \pm 0.6\%$. We note that this state fidelity rivals its state-of-the-art monolithic counterparts~\cite{Barends2014,Song2017,Wei2020,Marques2021}, highlighting the capability and potential of the low-loss interconnects. We then perform two more CNOT gates in parallel, yielding a 6-qubit GHZ state $|\psi_{6}\rangle = (|0\rangle^{\otimes 6}+|1\rangle^{\otimes 6})\sqrt{2}$, see Fig.~\ref{fig4}{\bf c}, with a fidelity of $\mathcal{F}_{6}=83.2\pm 0.8\%$.
To further expand the entanglement, we distribute the GHZ state to module $E$ using the 10$^{\rm th}$ mode and apply more CNOT gates, eventually yielding a 10-qubit (step III) and 12-qubit GHZ state (step IV), where module $E$ is chosen because of its higher performance interconnect; see Supplementary Information.
As quantum state tomography requires measurements and computational resources that grow exponentially with the number of qubits, full characterization of these two states is impractical~\cite{Song2019}. Nevertheless, the density matrix $\rho$ of an ideal $N$-qubit GHZ state has only four nonzero elements: two diagonal elements $\rho_{0...0,0...0}$ and $\rho_{1...1,1...1}$, and two conjugate off-diagonal elements $\rho_{0...0,1...1}$ and $\rho_{1...1,0...0}$, allowing one to evaluate the state fidelity in an easier way. The diagonal elements are simply the qubit joint-readout probabilities $P_{0...0}$ and $P_{1...1}$. To access the off-diagonal terms, we measure the expectation value of the operator $\mathcal{P}(\gamma) =\otimes_{q=1}^N (\cos\gamma X_q + \sin\gamma Y_q)$~\cite{Song2019}, where $q$ is the qubit index, $X_q$, $Y_q$ the Pauli operators for qubit $q$, and $\gamma\in [0,2\pi]$. Ideally, $\langle \mathcal{P}(\gamma) \rangle = 2|\rho_{1...1,0...0}|\cos(N\gamma+\phi_0)$, where $\phi_0$ is the phase of $\rho_{1...1,0...0}$. The oscillation period of $2\pi/N$ is a unique feature of the $N$-qubit GHZ state. The GHZ state fidelity is then given by $\mathcal{F}_{N}=(P_{0...0}+P_{1...1}+2|\rho_{1...1,0...0}|)/2$. In Fig.~\ref{fig4}{\bf d} and {\bf e}, we show the measured $\langle \mathcal{P} (\gamma)\rangle$ for the 10-qubit and 12-qubit GHZ states respectively. Combining these measurements with the qubit joint-readout probabilities, we estimate the GHZ state fidelities $\mathcal{F}_{10}=63.0\pm 1.5\%$ and $\mathcal{F}_{12}=55.8\pm 1.8\%$ respectively, clearly above the threshold of 1/2 for genuine multipartite entanglement~\cite{Guhne2010}. To our knowledge, this represents the largest quantum entanglement generated in solid-state systems separated by a macroscopic distance to date.  We are not able to entangle all the qubits on this modular processor with a meaningful fidelity,  limited mainly by qubit decoherence and the CZ gate error.

In summary, we have demonstrated low-loss quantum interconnects that reach the native coherence of superconducting qubits.
With linear loss as low as optical fibers, these cables can path new avenues for a broad spectrum of microwave applications at cryogenic temperatures, including modular quantum computing, quantum networks~\cite{Kimble2008}, quantum optics~\cite{Blais2020}, quantum nonlocality test~\cite{Hensen2015}, hybrid quantum systems~\cite{Tabuchi2015}, microwave delay lines~\cite{Su2008}, and radio astronomy.
In particular, we have demonstrated a path to building large-scale modular quantum computers, using high-performance modules linked by these low-loss interconnects.
We have shown high-fidelity inter-module quantum state transfer, entanglement generation, and large-scale entangled states linking multiple quantum modules, rivaling that of monolithic designs.
With these demonstrations, modular superconducting quantum processors are now poised to scale up to large size and explore more sophisticated quantum computation and simulation tasks.

\clearpage

\setlength{\parskip}{12pt}%

\clearpage

\bigskip
\bibliographystyle{naturemag}
\bibliography{bibliography}

\subsection*{Funding Information}
This work was supported by the Key-Area Research and Development Program of Guangdong Province (2018B030326001), the National Natural Science Foundation of China (U1801661, 12174178), the Guangdong Innovative and Entrepreneurial Research Team Program (2016ZT06D348), the Guangdong Provincial Key Laboratory (2019B121203002), the Science, Technology and Innovation Commission of Shenzhen Municipality (KYTDPT20181011104202253, KQTD20210811090049034), the Shenzhen-Hong Kong Cooperation Zone for Technology and Innovation (HZQB-KCZYB-2020050), and the NSF of Beijing (Z190012).

\subsection*{Acknowledgements}
We thank Yu He and Junqiu Liu for helpful discussions and critical reading of the manuscript.

\subsection*{Author contributions}
Yo.Z. conceived the idea and supervised the experiment. S.L. supervised the device fabrication and supported the infrastructure setup. J.N. and Yo.Z. wirebonded the cables, performed the measurement and analyzed the data. L.Z. fabricated the devices and designed the sample holder. Y.L. and J.Q. assisted the measurement.
Yo.Z. built the custom microwave electronics. Yo.Z. and A.N.C wrote the manuscript. All authors contributed to discussions and production of the manuscript.

\subsection*{Additional information}
Correspondence and requests for materials should be addressed to Yo.Z and S.L. \\
(\texttt{zhongyp@sustech.edu.cn, lius3@sustech.edu.cn})

\subsection*{Competing financial interests}
Yo.Z., S.L., D.Y., J.N. and L.Z. are inventors on a provisional patent application that has been filed relating to this work. Other authors declare no competing interests.

\end{document}


\title{Supplementary Information for ``Low-loss interconnects for modular superconducting quantum processors''}
\author{Jingjing Niu}
\affiliation{Shenzhen Institute for Quantum Science and Engineering, Southern University of Science and Technology, Shenzhen 518055, China}
\affiliation{International Quantum Academy, Shenzhen 518048, China}
\affiliation{Guangdong Provincial Key Laboratory of Quantum Science and Engineering, Southern University of Science and Technology, Shenzhen 518055, China}
\author{Libo Zhang}
\affiliation{Shenzhen Institute for Quantum Science and Engineering, Southern University of Science and Technology, Shenzhen 518055, China}
\affiliation{International Quantum Academy, Shenzhen 518048, China}
\affiliation{Guangdong Provincial Key Laboratory of Quantum Science and Engineering, Southern University of Science and Technology, Shenzhen 518055, China}
\author{Yang Liu}
\affiliation{Shenzhen Institute for Quantum Science and Engineering, Southern University of Science and Technology, Shenzhen 518055, China}
\affiliation{International Quantum Academy, Shenzhen 518048, China}
\affiliation{Guangdong Provincial Key Laboratory of Quantum Science and Engineering, Southern University of Science and Technology, Shenzhen 518055, China}
\author{Jiawei Qiu}
\affiliation{Shenzhen Institute for Quantum Science and Engineering, Southern University of Science and Technology, Shenzhen 518055, China}
\affiliation{International Quantum Academy, Shenzhen 518048, China}
\affiliation{Guangdong Provincial Key Laboratory of Quantum Science and Engineering, Southern University of Science and Technology, Shenzhen 518055, China}
\author{Wenhui Huang}
\affiliation{Shenzhen Institute for Quantum Science and Engineering, Southern University of Science and Technology, Shenzhen 518055, China}
\affiliation{International Quantum Academy, Shenzhen 518048, China}
\affiliation{Guangdong Provincial Key Laboratory of Quantum Science and Engineering, Southern University of Science and Technology, Shenzhen 518055, China}
\author{Jiaxiang Huang}
\affiliation{Shenzhen Institute for Quantum Science and Engineering, Southern University of Science and Technology, Shenzhen 518055, China}
\affiliation{International Quantum Academy, Shenzhen 518048, China}
\affiliation{Guangdong Provincial Key Laboratory of Quantum Science and Engineering, Southern University of Science and Technology, Shenzhen 518055, China}
\author{Hao Jia}
\affiliation{Shenzhen Institute for Quantum Science and Engineering, Southern University of Science and Technology, Shenzhen 518055, China}
\affiliation{International Quantum Academy, Shenzhen 518048, China}
\affiliation{Guangdong Provincial Key Laboratory of Quantum Science and Engineering, Southern University of Science and Technology, Shenzhen 518055, China}
\author{Jiawei Liu}
\affiliation{Shenzhen Institute for Quantum Science and Engineering, Southern University of Science and Technology, Shenzhen 518055, China}
\affiliation{International Quantum Academy, Shenzhen 518048, China}
\affiliation{Guangdong Provincial Key Laboratory of Quantum Science and Engineering, Southern University of Science and Technology, Shenzhen 518055, China}
\author{Ziyu Tao}
\affiliation{Shenzhen Institute for Quantum Science and Engineering, Southern University of Science and Technology, Shenzhen 518055, China}
\affiliation{International Quantum Academy, Shenzhen 518048, China}
\affiliation{Guangdong Provincial Key Laboratory of Quantum Science and Engineering, Southern University of Science and Technology, Shenzhen 518055, China}
\author{Weiwei Wei}
\affiliation{Shenzhen Institute for Quantum Science and Engineering, Southern University of Science and Technology, Shenzhen 518055, China}
\affiliation{International Quantum Academy, Shenzhen 518048, China}
\affiliation{Guangdong Provincial Key Laboratory of Quantum Science and Engineering, Southern University of Science and Technology, Shenzhen 518055, China}
\author{Yuxuan Zhou}
\affiliation{Shenzhen Institute for Quantum Science and Engineering, Southern University of Science and Technology, Shenzhen 518055, China}
\affiliation{International Quantum Academy, Shenzhen 518048, China}
\affiliation{Guangdong Provincial Key Laboratory of Quantum Science and Engineering, Southern University of Science and Technology, Shenzhen 518055, China}
\author{Wanjing Zou}
\affiliation{Shenzhen Institute for Quantum Science and Engineering, Southern University of Science and Technology, Shenzhen 518055, China}
\affiliation{International Quantum Academy, Shenzhen 518048, China}
\affiliation{Guangdong Provincial Key Laboratory of Quantum Science and Engineering, Southern University of Science and Technology, Shenzhen 518055, China}
\author{Yuanzhen Chen}
\affiliation{Shenzhen Institute for Quantum Science and Engineering, Southern University of Science and Technology, Shenzhen 518055, China}
\affiliation{International Quantum Academy, Shenzhen 518048, China}
\affiliation{Guangdong Provincial Key Laboratory of Quantum Science and Engineering, Southern University of Science and Technology, Shenzhen 518055, China}
\affiliation{Department of Physics, Southern University of Science and Technology, Shenzhen 518055, China}
\author{Xiaowei Deng}
\affiliation{Shenzhen Institute for Quantum Science and Engineering, Southern University of Science and Technology, Shenzhen 518055, China}
\affiliation{International Quantum Academy, Shenzhen 518048, China}
\affiliation{Guangdong Provincial Key Laboratory of Quantum Science and Engineering, Southern University of Science and Technology, Shenzhen 518055, China}
\author{Xiuhao Deng}
\affiliation{Shenzhen Institute for Quantum Science and Engineering, Southern University of Science and Technology, Shenzhen 518055, China}
\affiliation{International Quantum Academy, Shenzhen 518048, China}
\affiliation{Guangdong Provincial Key Laboratory of Quantum Science and Engineering, Southern University of Science and Technology, Shenzhen 518055, China}
\author{Changkang Hu}
\affiliation{Shenzhen Institute for Quantum Science and Engineering, Southern University of Science and Technology, Shenzhen 518055, China}
\affiliation{International Quantum Academy, Shenzhen 518048, China}
\affiliation{Guangdong Provincial Key Laboratory of Quantum Science and Engineering, Southern University of Science and Technology, Shenzhen 518055, China}
\author{Ling Hu}
\affiliation{Shenzhen Institute for Quantum Science and Engineering, Southern University of Science and Technology, Shenzhen 518055, China}
\affiliation{International Quantum Academy, Shenzhen 518048, China}
\affiliation{Guangdong Provincial Key Laboratory of Quantum Science and Engineering, Southern University of Science and Technology, Shenzhen 518055, China}
\author{Jian Li}
\affiliation{Shenzhen Institute for Quantum Science and Engineering, Southern University of Science and Technology, Shenzhen 518055, China}
\affiliation{International Quantum Academy, Shenzhen 518048, China}
\affiliation{Guangdong Provincial Key Laboratory of Quantum Science and Engineering, Southern University of Science and Technology, Shenzhen 518055, China}
\author{Dian Tan}
\affiliation{Shenzhen Institute for Quantum Science and Engineering, Southern University of Science and Technology, Shenzhen 518055, China}
\affiliation{International Quantum Academy, Shenzhen 518048, China}
\affiliation{Guangdong Provincial Key Laboratory of Quantum Science and Engineering, Southern University of Science and Technology, Shenzhen 518055, China}
\author{Yuan Xu}
\affiliation{Shenzhen Institute for Quantum Science and Engineering, Southern University of Science and Technology, Shenzhen 518055, China}
\affiliation{International Quantum Academy, Shenzhen 518048, China}
\affiliation{Guangdong Provincial Key Laboratory of Quantum Science and Engineering, Southern University of Science and Technology, Shenzhen 518055, China}
\author{Fei Yan}
\affiliation{Shenzhen Institute for Quantum Science and Engineering, Southern University of Science and Technology, Shenzhen 518055, China}
\affiliation{International Quantum Academy, Shenzhen 518048, China}
\affiliation{Guangdong Provincial Key Laboratory of Quantum Science and Engineering, Southern University of Science and Technology, Shenzhen 518055, China}
\author{Tongxing Yan}
\affiliation{Shenzhen Institute for Quantum Science and Engineering, Southern University of Science and Technology, Shenzhen 518055, China}
\affiliation{International Quantum Academy, Shenzhen 518048, China}
\affiliation{Guangdong Provincial Key Laboratory of Quantum Science and Engineering, Southern University of Science and Technology, Shenzhen 518055, China}
\author{Song Liu}
\email{lius3@sustech.edu.cn}
\affiliation{Shenzhen Institute for Quantum Science and Engineering, Southern University of Science and Technology, Shenzhen 518055, China}
\affiliation{International Quantum Academy, Shenzhen 518048, China}
\affiliation{Guangdong Provincial Key Laboratory of Quantum Science and Engineering, Southern University of Science and Technology, Shenzhen 518055, China}
\author{Youpeng Zhong}
\email{zhongyp@sustech.edu.cn}
\affiliation{Shenzhen Institute for Quantum Science and Engineering, Southern University of Science and Technology, Shenzhen 518055, China}
\affiliation{International Quantum Academy, Shenzhen 518048, China}
\affiliation{Guangdong Provincial Key Laboratory of Quantum Science and Engineering, Southern University of Science and Technology, Shenzhen 518055, China}
\author{Andrew N. Cleland}
\affiliation{Pritzker School of Molecular Engineering, University of Chicago, Chicago IL 60637, USA}
\affiliation{Center for Molecular Engineering and Material Science Division, Argonne National Laboratory, Argonne IL 60439, USA}
\author{Dapeng Yu}
\affiliation{Shenzhen Institute for Quantum Science and Engineering, Southern University of Science and Technology, Shenzhen 518055, China}
\affiliation{International Quantum Academy, Shenzhen 518048, China}
\affiliation{Guangdong Provincial Key Laboratory of Quantum Science and Engineering, Southern University of Science and Technology, Shenzhen 518055, China}
\affiliation{Department of Physics, Southern University of Science and Technology, Shenzhen 518055, China}

\maketitle

\setcounter{equation}{0}
\setcounter{figure}{0}
\setcounter{table}{0}
\setcounter{page}{1}

\renewcommand{\theequation}{S\arabic{equation}}
\renewcommand{\thefigure}{S\arabic{figure}}
\renewcommand{\thetable}{S\arabic{table}}

\section{Characterization of superconducting cables}\label{sec:wirebond}
The low-loss pure aluminum coaxial cables used in this experiment have an outer conductor diameter of 2.1 mm, an inner conductor diameter of 0.54 mm, and a dielectric insulating layer made of low-density polytetrafluoroethylene (PTFE). Similar to Ref.~\onlinecite{Zhong2021}, we avoid the use of normal-metal connectors (e.g. SMA connectors) and instead use $25~\mu$m diameter aluminum wires to bond the cables directly to the superconducting quantum chips. The larger cable size here allows us to bond multiple wires to the inner conductor, whereas in Ref.~\onlinecite{Zhong2021} the 0.2 mm diameter inner conductor could barely allow a single wirebond connection. In addition, the thin native oxide layer naturally forming on the aluminum surface can be removed by ultrasonic bonding~\cite{Long2018}. These properties make the all-aluminum connection between the quantum processor and the cables more reliable and less lossy.

\begin{figure}[H]
  \centering
  \includegraphics[width=0.8\textwidth]{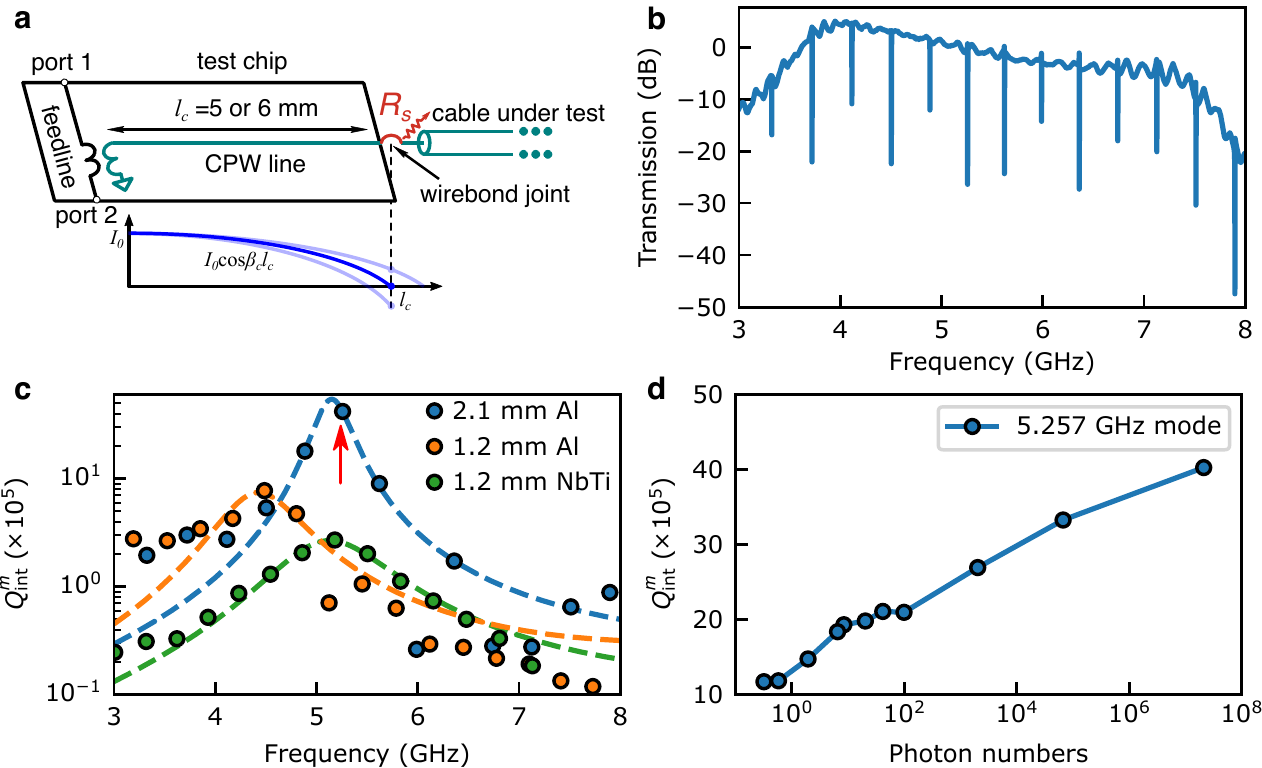}
  \caption{\label{cable}
  {\bf Characterization of superconducting cables.}
  {\bf a,} Schematic of the measurement setup, where on the test chip, a CPW line is inductively coupled to a feedline in a hanger geometry at the shorted end, and wirebonded to the cable under test at the open end.
  {\bf b,} Transmission measurement of a 2.1 mm Al coaxial cable using a vector network analyzer, where a series of dips correspond to the standing wave modes.
  {\bf c,} The $Q_{\rm int}^m$ of 3 types of cables obtained from the transmission measurement in {\bf b}. The 2.1 mm Al cable and 1.2 mm NbTi cable are connected to $\ell_c=5$~mm CPW lines, and the 1.2 mm Al cable is connected to a $\ell_c=6$~mm CPW line.
  {\bf d,} The power dependent $Q_{\rm int}^m$ of the 2.1 mm Al cable mode at 5.257~GHz (marked by an arrow in {\bf c}), with $Q_{\rm int}^m=1.2\times 10^6$ at single photon levels.
}
\end{figure}

 Prior to the assembly of the modular quantum processor, we have screened several kinds of superconducting coaxial cables at cryogenic temperatures independently using transmission measurement with dedicated test chips, where the 2.1 mm Al cable is found to have best performance among the tested ones, see Fig.~\ref{cable}. On the test chip, a coplanar waveguide (CPW) line of length $\ell_c=5$ or $6$~mm is inductively coupled to a feedline in a hanger geometry at the shorted end for transmission measurement, and wirebonded to the cable under test at the open end, see Fig.~\ref{cable}{\bf a}. The other end of the cable is left open. The loss in the wirebond joint is modeled as a series resistance $R_s$. For simplicity, we assume the current profile of the standing wave modes follows a simple cosine shape along the $\ell_c=5$ or 6~mm CPW line, which is essentially a quarter-wavelength impedance transformer~\cite{Pozar} here converting the standing wave mode current at the wirebond joint from $I_0$ to $I_0\cos(\beta_c\ell_c)$~\cite{Pozar}, where $I_0$ is the maximum current amplitude at the shorted end, and $\beta_c$ is the propagation constant of the CPW line (see Fig.~\ref{cable}{\bf a}). This current creates a loss power of $P_{\rm loss} = I_0^2 \cos^2(\beta_c\ell_c) R_s$, loading the mode with a quality factor of~\cite{Pozar}
\begin{equation}\label{Qloss}
   Q_{\rm loss} = \omega_m \frac{L_m I_0^2}{P_{\rm loss}} = \omega_m \frac{L_m}{\cos^2(\beta_c\ell_c)R_s},
\end{equation}
here $m$ is the mode number, $L_m$ is the equivalent lumped element inductance of the standing wave mode, see section~\ref{sec:mmode}. The internal quality factor $Q_{\rm int}^{m}$ of the $m$-th standing wave mode is then given by
\begin{equation}\label{Qtotal}
    1/Q_{\rm int}^{m} = 1/ Q_{\rm loss} + 1/Q_{cb},
\end{equation}
where $Q_{cb}$ is the cable's intrinsic quality factor. For simplicity we have assumed that $Q_{cb}$ and $R_s$ are frequency independent, and the on-chip CPW line has the same intrinsic quality factor as the cable here.
When $\ell_c$ matches the quarter-wavelength of a standing wave mode, $\cos(\beta_c\ell_c)\approx 0$, therefore the dissipation through $R_s$ vanishes, and $Q_{\rm int}^{m}\approx Q_{cb}$.

From the transmission measurement (see Fig.~\ref{cable}{\bf b}) of the feedline using a vector network analyzer, we can fit the internal quality factor $Q_{\rm int}^m$ of the $m$-th standing wave mode~\cite{Megrant2012}. Three types of coaxial cables have been tested: (a) pure Al cables with 2.1 mm diameter, (b) pure Al cables with 1.2 mm diameter, and (c) NbTi cables with 1.2 mm diameter, see Fig.~\ref{cable}{\bf c}. Both Al cables are custom ordered from Hermerc Inc., whereas the NbTi cable is ordered from Keycom Corp. The 2.1 mm Al cable and 1.2 mm NbTi cable are connected to $\ell_c=5$~mm CPW lines during the test, whereas the 1.2 mm Al cable is connected to a $\ell_c=6$~mm CPW line. The dashed lines in Fig.~\ref{cable}{\bf c} are fits to Eqs.~\ref{Qloss} and~\ref{Qtotal}.

 It is found that the 1.2~mm NbTi cable has an intrinsic $Q_{cb}=2.7\times 10^5$, similar to the findings in Refs.~\onlinecite{Kurpiers2017,Burkhart2021,Zhong2021}, where Ref.~\onlinecite{Kurpiers2017} reported a $Q_{cb}$ as high as $0.92 \times 10^5$ for a 2.2 mm diameter NbTi coaxial cable from Keycom Corp.; in Ref.~\onlinecite{Burkhart2021}, typical $Q_{\rm int}^m$ of order $0.5 \times 10^5$ with occasional values as high as $1.6 \times 10^5$ were observed with a 0.86 mm diameter NbTi cable from Coax Co., suggesting $Q_{cb}>1.6\times 10^5$; and Ref.~\onlinecite{Zhong2021} estimated $Q_{cb}=0.91\times 10^5$ for the same kind of NbTi cable used in Ref.~\onlinecite{Burkhart2021}. See Table~\ref{cableCompare} for comparison. It is worth mentioning that long CPW delay lines fabricated on sapphire substrates with air-bridge crossovers show similar performance as the NbTi cables~\cite{Zhong2019,Chang2020}.

\begin{table}[H]
\begin{center}
\begin{tabular}{|c |c | c| c |c |c |c |}
  \hline
  \hline
  source                   & this work             & this work      & this work         & Ref.~\onlinecite{Zhong2021}  & Ref.~\onlinecite{Burkhart2021} & Ref.~\onlinecite{Kurpiers2017}\\
  \hline
  conductor material       & Al                    & Al             & NbTi               & NbTi                         & NbTi & NbTi\\
  cable diameter           & 2.1~mm                & 1.2~mm         & 1.2~mm            & 0.86~mm                      & 0.86~mm & 2.2~mm\\
  intrinsic $Q_{cb}$ & $4.2\times 10^6$      & $7.4\times 10^5$ & $2.7\times 10^5$  & $0.91\times 10^5$            & $1.6\times 10^5$ & $0.92\times 10^5$\\
  \hline
  \hline
\end{tabular}
\end{center}
\caption{\label{cableCompare} {\bf Comparison of superconducting cables.} All these coaxial cables use PTFE for insulation. Note the $Q_{cb}$ data from different sources are not obtained with the same standard, therefore are not suitable for quantitative comparison, but qualitatively we can clearly see that the Al cables are significantly better than the NbTi ones. Affected by TLS defects, the $Q_{cb}$ of the 2.1~mm Al cable drops at low power and reaches $1.2\times 10^6$ at single photon levels.}
\end{table}

  As shown in Fig.~\ref{cable}{\bf c}, the pure Al cables with 2.1 mm diameter has best performance, with a quality factor as high as $4.2\times 10^6$ at high power, corresponding to a linear loss of 0.15 dB/km. This loss rivals the value of 0.2 dB/km for optical fibers.
 In Fig.~\ref{cable}{\bf d}, we carry out power dependent measurement on the standing wave mode with the highest $Q_{\rm int}^m$ (marked by an arrow in Fig.~\ref{cable}{\bf c}), and find that its $Q_{\rm int}^m$ drops when the input power decreases, as affected by two-level-system (TLS) defects, reaching $Q_{\rm int}^m=1.2\times 10^6$ at single photon levels. This performance rivals the state-of-the-art CPW resonators fabricated on single crystal sapphire substrate using ultra-high vacuum molecular beam epitaxy (MBE) system~\cite{Megrant2012}, suggesting that the loss tangent of the low-density PTFE is $<10^{-6}$ in the low temperature, single-photon regime. This result is indeed surprising when compared with the routinely used commercial NbTi cables (see Table~\ref{cableCompare}). Nevertheless, a loss tangent of $2.3\times 10^{-6}$ for bulk PTFE at 28~K and 18.92~GHz has been reported in Ref.~\onlinecite{Jacob2002}. The loss tangent here is smaller than that in Ref.~\onlinecite{Jacob2002}, likely because the PTFE is low-density, and the temperature and frequency are lower compared to Ref.~\onlinecite{Jacob2002}. This suggests that the quality of the commercial NbTi cables at cryogenic temperatures of $\sim$10~mK is not limited by the PTFE dielectric, but rather by the conductor.

\begin{figure}[H]
\begin{center}
	\includegraphics[width=0.8\textwidth]{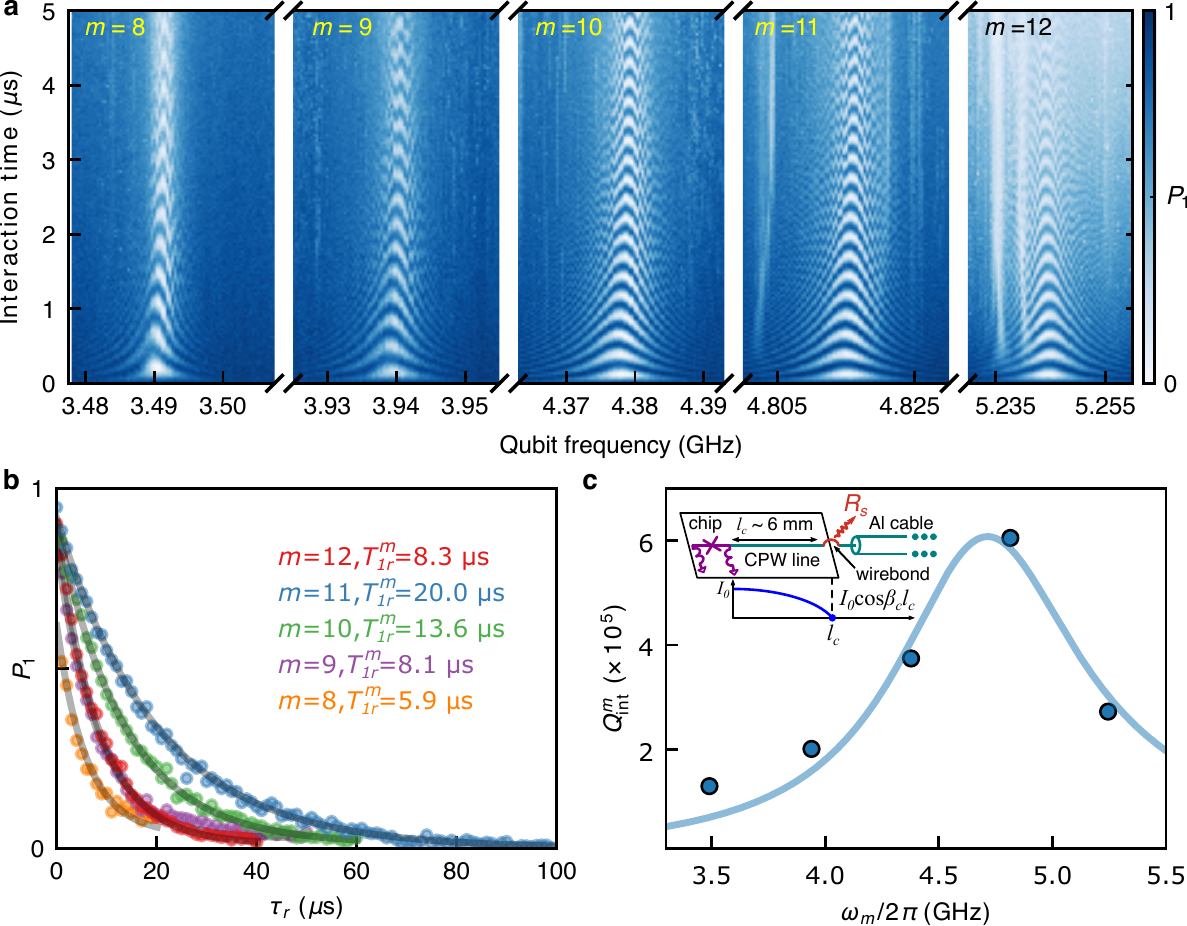}
	\caption{
    \label{fig2}
    {\bf Five standing wave modes in another modular processor assembly.}
    {\bf a,} In another modular processor assembly, the qubits have a larger frequency tuning range, allowing access to five standing modes. Here shows the vacuum Rabi oscillation between $Q_1^A$ and the $m=8$--$12$ standing wave modes in the $A$-$B$ interconnect.
    {\bf b,} $T_{1r}^{m}$ measurement of each standing wave mode. The $m=8$--12 modes have a lifetime of $T_{1r}^m=5.9$~$\mu$s, $8.1$~$\mu$s, $13.6$~$\mu$s, $20.0$~$\mu$s and $8.3$~$\mu$s respectively.
    {\bf c,} The $Q_{\rm int}^{m}$ of each mode versus the mode frequency $\omega_m/2\pi$. The line is fitting to Eqs.~\ref{Qloss} and~\ref{Qtotal}, with $Q_{cb}=6.0\times 10^5$. Inset: the effect of the CPW impedance transformer.
    }
\end{center}
\end{figure}
In the main text, limited by the narrow qubit frequency tuning range of $\sim 4.2$~GHz to $\sim 5.1$~GHz determined by the asymmetry of the qubit junctions, we can access the $m=10$ and 11 standing wave modes only. In addition to the assembly in the main text, we have another modular processor assembly where the qubits have a larger frequency tuning range of $\sim 3.3$~GHz to $\sim 5.5$~GHz, allowing for access to five standing wave modes, and the $Q_{\rm int}^{m}$ versus $\omega_m$ agrees well with Eqs.~\ref{Qloss} and \ref{Qtotal} (note the $R_s$ in Eq.~\ref{Qloss} should be replaced by $2R_s$ because there are two wirebond connections here), see Fig.~\ref{fig2}. In this assembly, the on-chip CPW line length $\ell_c=6$~mm is chosen to match the $m=11$ mode at $\omega_m/2\pi=4.815$~GHz, whereas in the assembly described in the main text, $\ell_c$ is slightly increased to 6.5~mm to match the $m=10$ mode at $\omega_m/2\pi=4.45$~GHz.

In this experiment, the rigid Al cables are firmly clamped to the qubit aluminum boxes on both sides. The cable standing modes have 2-3\% residual population measured using a Rabi population measurement method~\cite{Geerlings2013,Satzinger2018}, similar to that of the qubits, suggesting the cables are well thermalized.

\section{Experimental setup}\label{sec:setup}
\begin{figure}[H]
  \centering
  \includegraphics[width=0.8\textwidth]{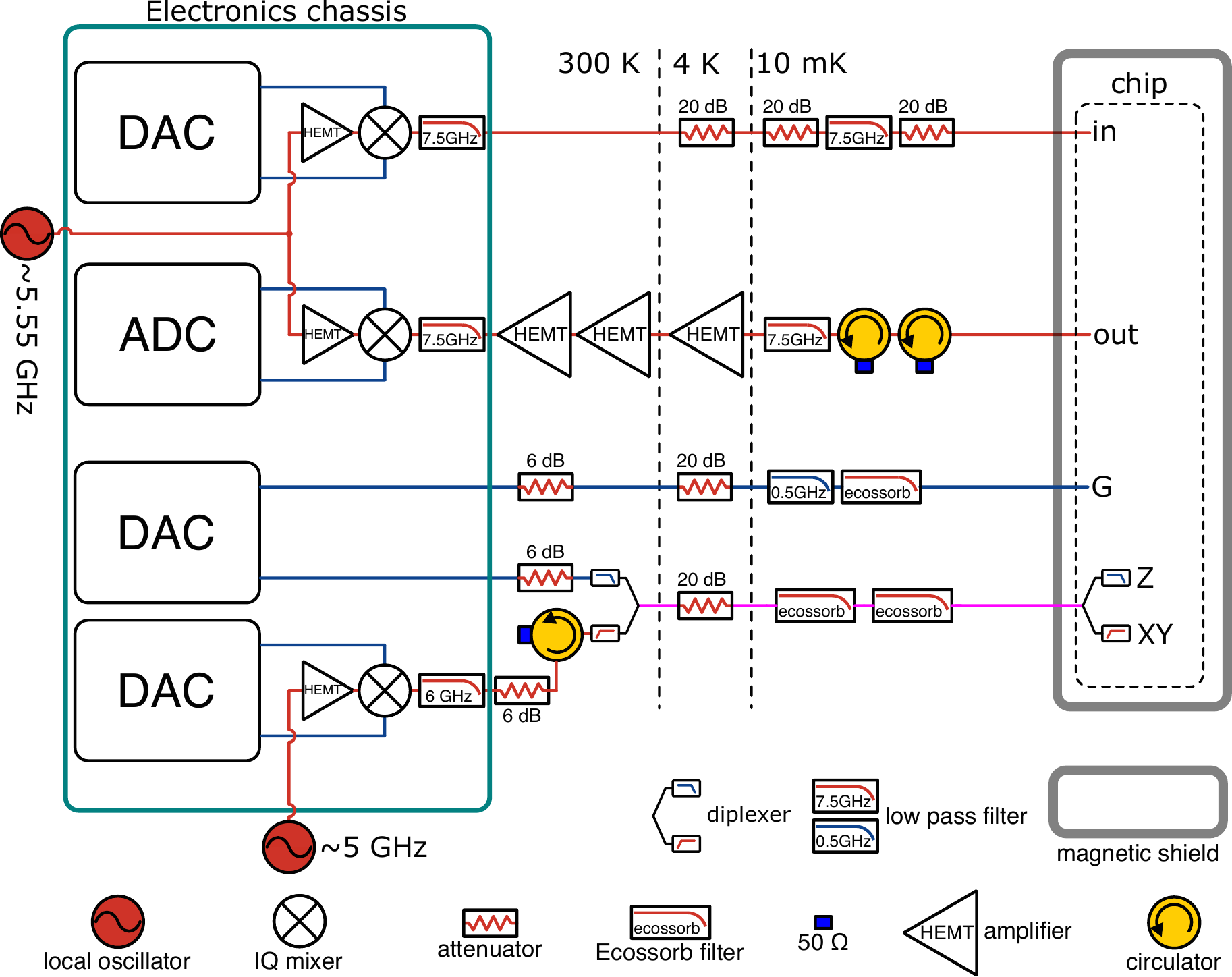}
  \caption{\label{setup}{\bf Control and readout wiring schematic.}}
\end{figure}

\begin{figure}[H]
  \centering
  \includegraphics[width=0.6\textwidth]{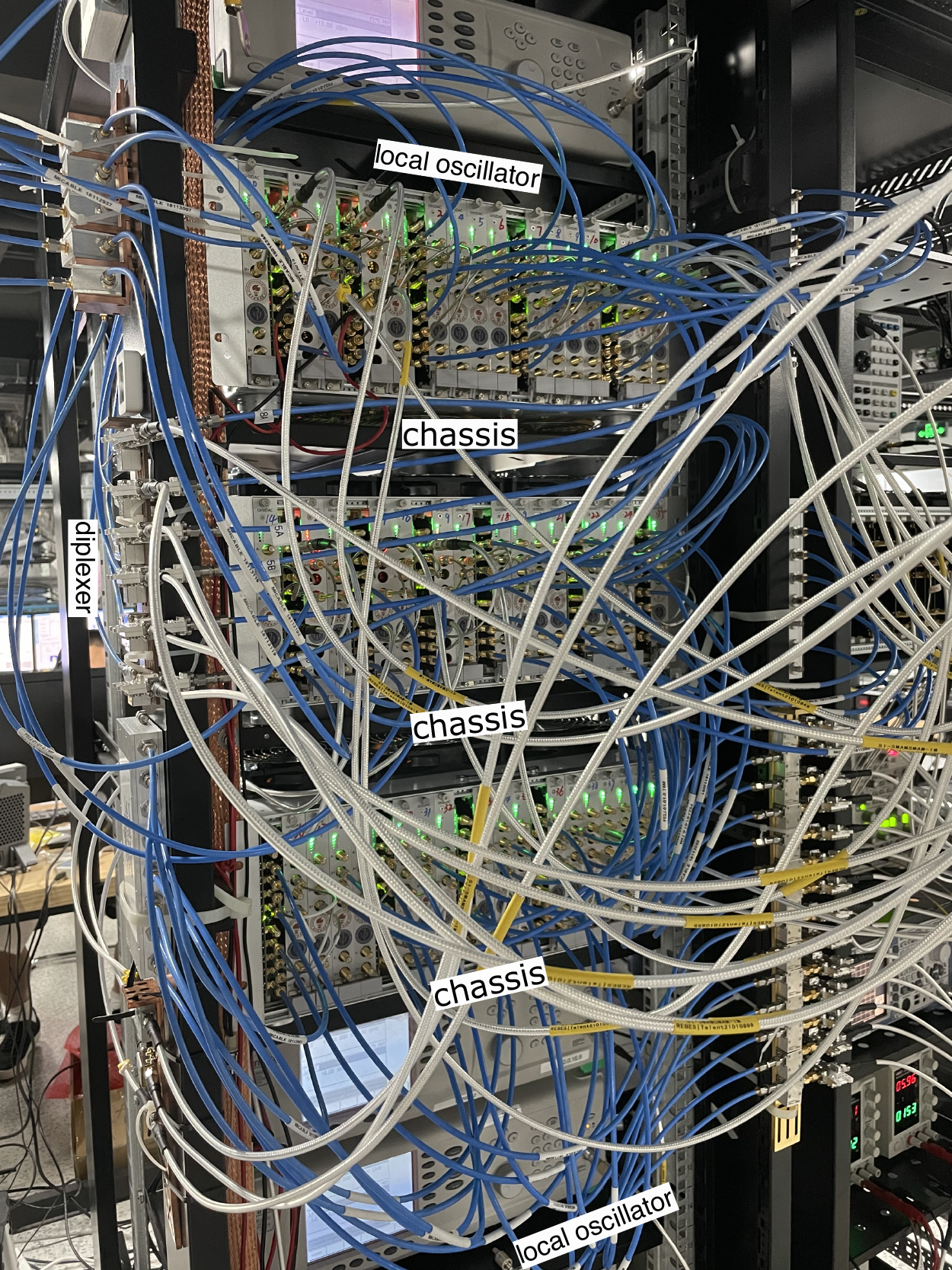}
  \caption{\label{chassis}
  {\bf The microwave electronics setup.}
  The electronics system used in this experiment consists of three 4U chassis and four Anritsu MG3692C local oscillators. Each chassis can host 21 DAC/ADC/mixer modules. Power, clock and trigger signals are routed to each module through the chassis backplane. The output of Anritsu MG3692C are divided with power splitters, then connected to the LO port of the mixer modules. }
\end{figure}

A representative schematic of the room-temperature electronics and the cryogenic wiring setup for qubit control and readout is shown in Fig.~\ref{setup}. We use custom digital-to-analog converter (DAC) and analog-to-digital converter (ADC) circuit board modules for qubit control and readout, respectively, both having a sampling rate of 1~Gs/s. The microwave electronics setup used in this experiment comprises 78 DAC channels, 10 ADC channels and 30 IQ mixers housed in three 4U chassis, and four Anritsu MG3692C local oscillators, see Fig.~\ref{chassis}.

In this experiment, each qubit requires 3 channels for qubit XY, Z and coupler (G) control respectively, adding up to 14 wiring channels including the input and output for each quantum module. Five modules thus require 70 channels, beyond the capacity of our BlueFors LD400 dilution fridge with normal density wiring. To save the cryogenic wiring resource, we combine the high frequency XY signal and the low frequency Z control signal using commercial broadband diplexers at room temperature, and deliver the combined signal to the qubit. At the mixing chamber stage with $\sim10$~mK base temperature, the combined signal is filtered with two Ecossorb filters from Hermerc Inc. which pass the low frequency Z signal, but attenuate the high frequency XY signal by $\sim 30$~dB. Inside the quantum module, an on-chip diplexer is implemented to separate the XY and Z signals before applying to the qubits, see the device design section~\ref{sec:design} for more details.

For the peripheral modules $B$ to $E$, 3 tunable couplers are not in use in this experiment. In such case, those unconnected quarter-wavelength CPW impedance transformers become resonators that can strongly couple to the qubits inadvertently. We can turn off the coupling by properly biasing the tunable couplers, but it costs extra control channels that are otherwise not used. To simplify the setup and save control channels, we disable these impedance transformers by shorting their open ends to ground using bonding wires. This changes their boundary conditions and converts them to short-ended half-wavelength resonators with resonant frequencies doubled, i.e., $\sim 10$~GHz, very far away from the qubit frequency and thus can be safely ignored. We can enable them again by carefully removing the shorting wires from the bonding pads if needed.

\section{Device characterization}
\subsection{Device design}\label{sec:design}

\begin{figure}[H]
  \centering
  \includegraphics[width=0.8\textwidth]{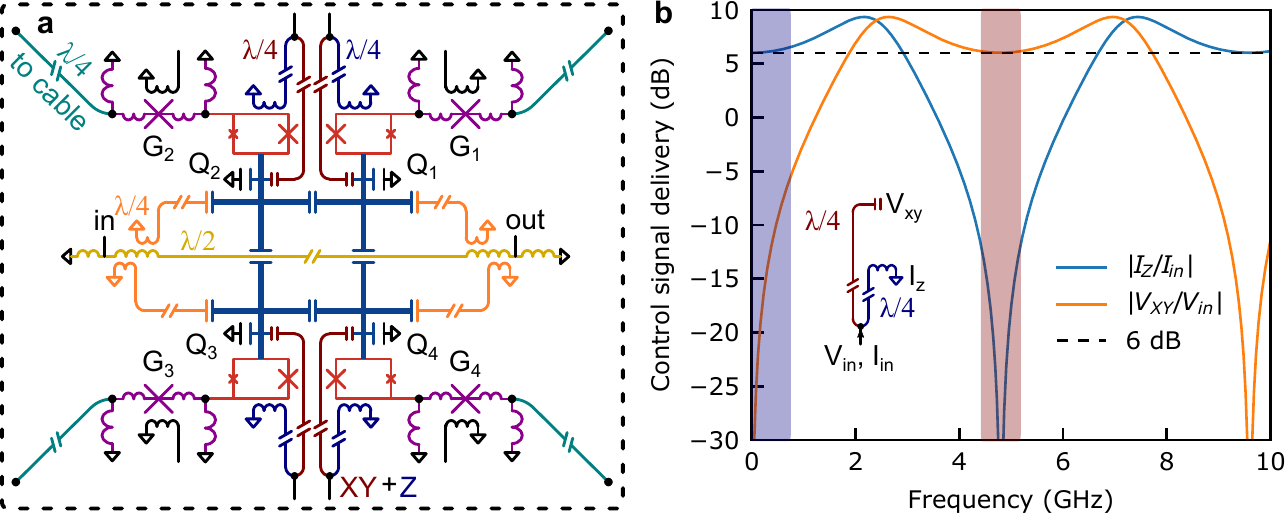}
  \caption{\label{device}
  {\bf a,} Detailed schematic of a quantum module. {\bf b,} Numerical simulation of control signal delivery for the on-chip diplexer. The low frequency bias signal is delivered to the Z arm (shaded in blue), whereas the high frequency microwave signal is deflected to the XY arm (shaded in red).}
\end{figure}

The assembly schematic in Fig.~1{\bf b} in the main text is greatly simplified. A more detailed version is provided in Fig.~\ref{device}{\bf a}. The blue crosses are qubit capacitor pads, the red rectangles with crosses are qubit junction loops (we use asymmetric junctions here to suppress dephasing noise~\cite{Hutchings2017}). The orange quarter-wavelength ($\lambda/4$) lines are qubit readout resonators, each capacitively coupled to a qubit and inductively coupled to the Purcell filter (yellow). The Purcell filter is essentially a half-wavelength ($\lambda/2$) CPW resonator with both ends shorted~\cite{Jeffrey2014,Satzinger2018,Bienfait2019,Zhong2021}, and the input and output lines intersect with the Purcell filter near the shorted ends.
The Purcell filter has a center frequency of about 5.5 GHz, a weak coupling to the input port (coupling $Q_c\sim2000$) and a strong coupling to the output port (coupling $Q_c\sim25$).
Each gmon tunable coupler is formed by two linear inductors and a Josephson junction (purple)~\cite{Chen2014}. A quarter-wavelength ($\lambda/4$) CPW impedance transformer (cyan) is connected to each gmon tunable coupler for cable connection.
As mentioned in the main text and in section~\ref{sec:setup}, we combine the qubit XY and Z control channels using commercial broadband diplexers at room temperature, and separate them with an on-chip diplexer in each quantum module. The on-chip diplexer here is essentially a combination of two quarter-wavelength ($\lambda/4$) CPW lines (see Fig.~\ref{device}), one short-ended (blue) and one open-ended (red). The short-ended line delivers low frequency bias current $I_Z$ to the qubit junction loop for Z control, while the open-ended line delivers high frequency microwave voltage signal $V_{XY}$ to the qubit capacitor for XY control. In Fig.~\ref{device}{\bf b}, we show the numerical simulation of control signal delivery for this on-chip diplexer, where we assume an input signal with $V_{in}$ and $I_{in}$ is applied, then we calculate the ratio $|I_Z/I_{in}|$ (the blue line) and $|V_{XY}/V_{in}|$ (the orange line). If the XY and Z signals were delivered separately, we should have $|V_{XY}/V_{in}|=|I_Z/I_{in}|=2$ due to the superposition of the input and totally reflected signals, as marked by the 6~dB dashed line in Fig.~\ref{device}{\bf b}. From the simulation, we see that at low frequency (shaded in blue), the XY arm has high impedance and does not interfere the current delivery to the Z arm. Near the resonant frequency of the two $\lambda/4$ CPW lines (shaded in red), we have~\cite{Pozar}
\begin{equation}\label{Ztransform}
  Z_{\lambda/4}\approx Z_0^2/Z_L,
\end{equation}
where $Z_0$ is the characteristic impedance of the CPW lines, and $Z_L$ is the load impedance. Since $Z_L=0$ for the short-ended Z arm, it becomes high impedance, conversely, the XY arm becomes low impedance, therefore the microwave signal is deflected to the XY arm. This design has $\sim600$~MHz bandwidth centered near 4.8~GHz with 20~dB isolation between $|V_{XY}/V_{in}|$ and $|I_Z/I_{in}|$, significantly suppressing the inadvertent modulation of the qubit frequency when applying XY drive signals~\cite{Naik2017,Manenti2021}.
As shown in the next subsection~\ref{sec:qubit}, this approach meets the stringent
requirement of achieving high fidelity gate operations meanwhile preserving the qubit coherence.
Other approaches combining the qubit XY and Z control signal have also been demonstrated~\cite{Manenti2021,Chu2021}.

\subsection{Characterization of the qubits}\label{sec:qubit}
\begin{figure}[H]
  \centering
  \includegraphics[width=0.8\textwidth]{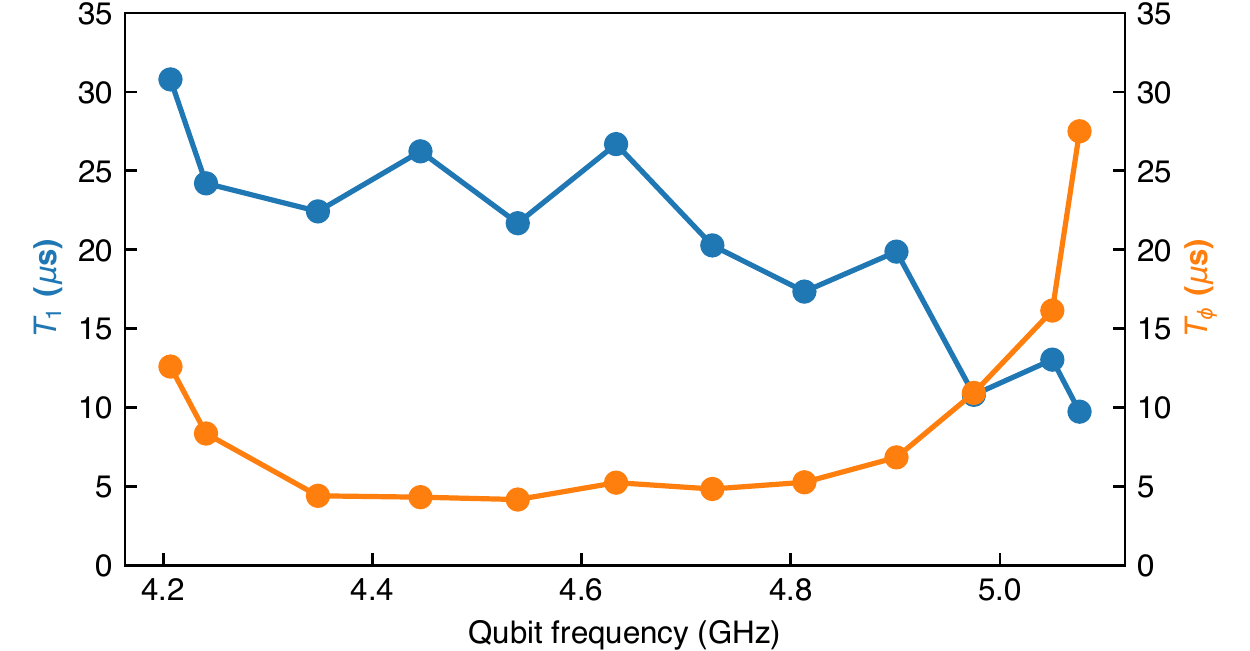}
  \caption{\label{t1t2}
  {\bf Typical qubit $T_1$ and $T_{\phi}$ versus qubit frequency.}
   With asymmetric junctions, the qubit frequency can be tuned from $\sim 4.2$~GHz to $\sim 5.1$~GHz. The decreasing trend of the qubit $T_1$ with frequency is due to Purcell decay as its frequency approaches the readout resonator frequency at $\sim5.5$~GHz. The qubit dephasing time $T_{\phi}$ is very long near the maximum and minimum sweet spots, as expected.}
\end{figure}
To suppress dephasing noise while retaining some frequency tunability, we use asymmetric Josephson junctions with $\alpha=E_{J1}/E_{J2}=5.3$, where $E_{J1}$ and $E_{J2}$ are the Josephson energies of the two qubit junctions~\cite{Hutchings2017}. This gives a qubit frequency tuning range of $\sim 4.2$~GHz to $\sim 5.1$~GHz. Figure~\ref{t1t2} shows the typical qubit lifetime $T_1$ and pure dephasing time $T_{\phi}$ at different frequencies.

The parameters and typical performance of each qubit are summarized in Table \ref{parameters}.
\begin{table}[H]
\begin{center}
\begin{tabular}{|l |c|c|c|c|c|c|c|}
  \hline
  \hline
  \rowcolor{gray}
   & $\omega_{10}/2\pi$ (GHz) & $\eta/2\pi$ (GHz)& $T_1$ ($\mu$s) & $T_\phi$ ($\mu$s) & $\omega_{rr}/2\pi$ (GHz) & $F_0$ & $F_1$  \\
  \hline
  $Q_1^A$ & 5.096 & $-$0.200 & 13.1 & 6.8  & 5.6968 & 0.948 & 0.920\\
  $Q_2^A$ & 4.344 & $-$0.221 & 35.5 & 2.7  & 5.5201 & 0.923 & 0.903\\
  $Q_3^A$ & 4.968 & $-$0.210 & 14.3 & 5.1  & 5.5745 & 0.911 & 0.884\\
  $Q_4^A$ & 4.520 & $-$0.219 & 18.4 & 2.8  & 5.6285 & 0.945 & 0.909\\
  \hline
  \rowcolor{lightgray}
  $Q_1^B$ & 4.852 & $-$0.209 & 21.2 & 5.5  & 5.5482 & 0.949 & 0.922\\
  \rowcolor{lightgray}
  $Q_2^B$ & 4.317 & $-$0.226 & 23.0 & 10.1  & 5.6216 & 0.942 & 0.918\\
  \rowcolor{lightgray}
  $Q_3^B$ & 5.096 & $-$0.211 & 19.3 & 6.4  & 5.6892 & 0.944 & 0.925\\
  \rowcolor{lightgray}
  $Q_4^B$ & 4.459 & $-$0.218 & 29.1 & 5.1  & 5.4959 & 0.909 & 0.888\\
  \hline
  $Q_1^C$ & 5.012 & $-$0.218 & 14.7 & 13.2  & 5.6835 & 0.966 & 0.937\\
  $Q_2^C$ & 4.657 & $-$0.206 & 15.6 & 4.7  & 5.4861 & 0.951 & 0.914\\
  $Q_3^C$ & 5.161 & $-$0.220 & 16.9 & 3.9  & 5.5383 & 0.981 & 0.935\\
  $Q_4^C$ & 4.534 & $-$0.214 & 21.7 & 3.0  & 5.6192 & 0.952 & 0.880\\
  \hline
  \rowcolor{lightgray}
  $Q_1^D$ & 4.561 & $-$0.196 & 14.4 & 6.2  & 5.6650 & 0.945 & 0.915\\
  \rowcolor{lightgray}
  $Q_2^D$ & 5.012 & $-$0.219 & 14.6 & 13.4  & 5.5165 & 0.915 & 0.897\\
  \rowcolor{lightgray}
  $Q_3^D$ & 4.495 & $-$0.210 & 25.4 & 5.1  & 5.5630 & 0.925 & 0.935\\
  \rowcolor{lightgray}
  $Q_4^D$ & 4.966 & $-$0.220 & 9.3  & 15.0  & 5.6035 & 0.979 & 0.925\\
  \hline
  $Q_1^E$ & 4.907 & $-$0.210 & 11.1 & 4.7  & 5.6259 & 0.905 & 0.898\\
  $Q_2^E$ & 4.544 & $-$0.200 & 27.1 & 4.6  & 5.6873 & 0.941 & 0.904\\
  $Q_3^E$ & 5.070 & $-$0.203 & 14.3 & 8.6  & 5.5176 & 0.908 & 0.855\\
  $Q_4^E$ & 4.237 & $-$0.215 & 30.6 & 7.9  & 5.5642 & 0.931 & 0.889\\
  \hline
  \hline
\end{tabular}
\end{center}
\caption{\label{parameters} {\bf Qubit parameters.} Here $\omega_{10}/2\pi$ is the qubit operating frequency, $\eta/2\pi$ is the qubit nonlinearity, $T_1$ and $T_\phi$ are the qubit lifetime and pure dephasing time at the operating frequency respectively, $\omega_{rr}/2\pi$ is the readout resonator frequency, $F_0$ and $F_1$ are the readout fidelities of the $|0\rangle$ and $|1\rangle$ states respectively.}
\end{table}

\begin{figure}[H]
  \centering
  \includegraphics[width=0.9\textwidth]{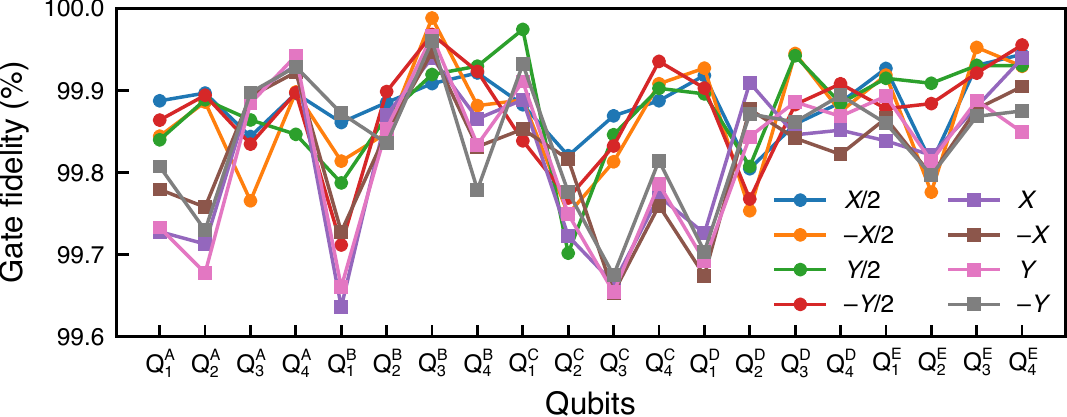}
  \caption{\label{RB_Fid} {\bf Single-qubit gates characterized with randomized benchmarking.} The average gate fidelity is 99.85\%.}
\end{figure}

We characterize the single-qubit gates using Clifford-based randomized benchmarking (RB)~\cite{Ryan2009,Brown2011,Magesan2012,Barends2014}. Figure~\ref{RB_Fid} summarizes the typical single-qubit gate fidelities for all qubits in this experiment, with an average fidelity of 99.85\%.

\subsection{Characterization of the CZ gates}\label{sec:cz}

\begin{figure}[H]
  \centering
  \includegraphics[width=0.9\textwidth]{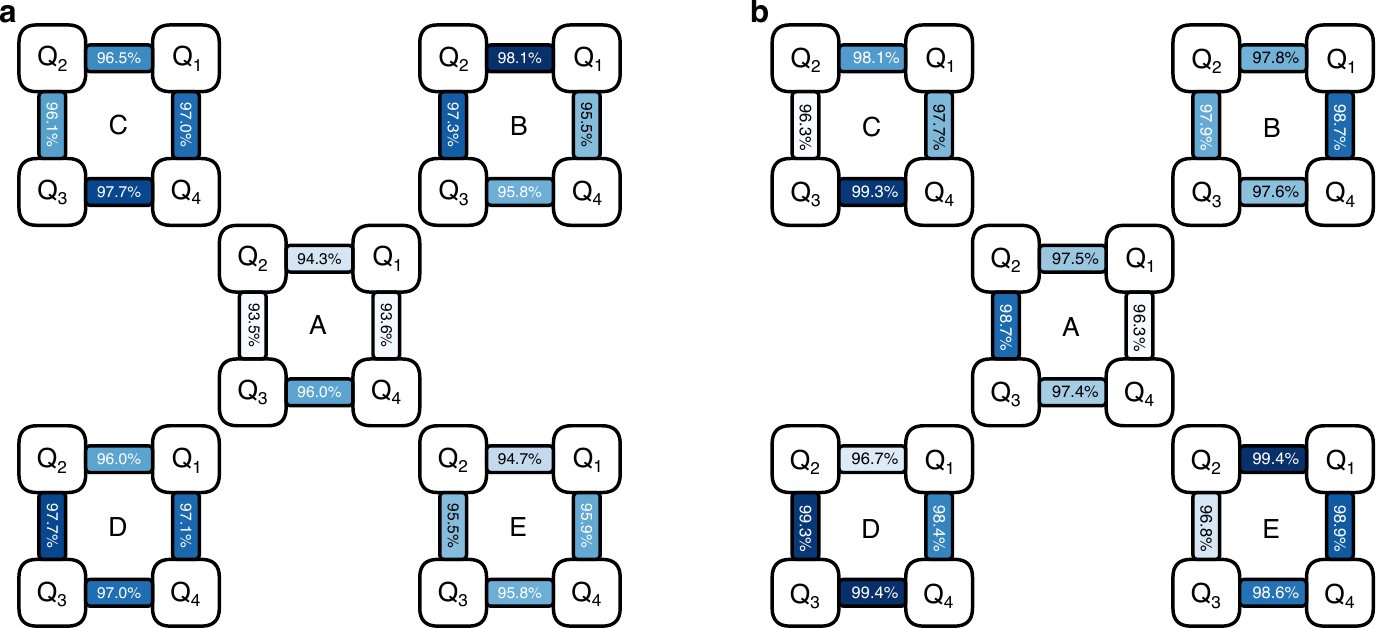}
  \caption{\label{CZ_Bell} {\bf Characterization of the CZ gates.}
  {\bf a,} CZ gate fidelities characterized by XEB, with an average fidelity of 96.1\%. The performance of module $A$ is slightly worse, likely caused by interference from the cable standing wave modes during gate operation.
  {\bf b,} Fidelities of the Bell states generated with the CZ gates.}
\end{figure}

In each module, the four qubits are capacitively coupled to each other with a fixed coupling strength of about 15~MHz. To suppress residual coupling, the adjacent qubits are detuned by a few hundred MHz when idling. To implement the CZ gate, we follow Ref.~\onlinecite{Barends2014} and tune the frequency of one qubit along a ``fast adiabatic'' trajectory which brings the $|11\rangle$ state close to the avoided-level crossing with the $|02\rangle$ state, such that a state-dependent relative phase shift of $\pi$ is accumulated. The CZ gate time here ranges from 40~ns to 45~ns.

We use the cross-entropy benchmarking (XEB) technique~\cite{Arute2019} to estimate the CZ gate fidelities, as summarized in Fig.~\ref{CZ_Bell}{\bf a}, with an average fidelity of 96.1\%. Randomized benchmarking (RB) has been routinely used to evaluate single-qubit gate fidelities as well as two-qubit gate fidelities~\cite{Ryan2009,Brown2011,Magesan2012,Barends2014}. However, two-qubit RB is significantly more complicated than single-qubit RB, involving the two-qubit Clifford group $C_2$ with 11,520 elements~\cite{Barends2014}. By repeatedly interleaving the target two-qubit gate with random single-qubit gates, XEB can estimate the two-qubit gate fidelity in a way that is less involved experimentally~\cite{Arute2019}.

The CZ gate fidelity characterizes its average performance over various input states. To further understand its performance for the specific task of GHZ state generation in the main text, we create the Bell triplet state by first preparing the control qubit in $(|0\rangle+|1\rangle)/\sqrt{2}$ superposition, then applying the CNOT gate comprising the CZ gate sandwiched by the $-Y/2$ and $Y/2$ gates of the target qubit, yielding a Bell triplet state $(|00\rangle+|11\rangle)/2$ in the end. The Bell state fidelities are shown in Fig.~\ref{CZ_Bell}{\bf b}, which are consistently higher than the corresponding CZ gate fidelities in Fig.~\ref{CZ_Bell}{\bf a}, suggesting the two-state leakage is a major error source when implementing the CZ gate.

We note that a dynamic phase is accumulated in each qubit during the two-qubit gates, due to the change of the qubit frequency during the interaction. This dynamic phase can be physically corrected by applying a calibrated Z rotation. Alternatively, here we adjust the phase of the second $Y/2$ gate on the target qubit to correct for the dynamic phase shift when performing a CNOT gate to simplify the control pulses.

\subsection{Qubit-cable coupling}\label{sec:mmode}
The 2.1~mm diameter Al cables have a specific capacitance $\mathscr{C}_{cb} = 86.5$~pF/m and a specific inductance $\mathscr{L}_{cb} = 216$~nH/m (provided by the cable manufacturer). In this experiment, each gmon tunable coupler is connected to a $\ell_{cb} = 0.25$~m long Al cable through a quarter-wavelength~($\lambda/4$) CPW line of length $\ell_c\approx6.5$~mm. The CPW line has a specific capacitance $\mathscr{C}_{cpw} = 173$~pF/m and a specific inductance $\mathscr{L}_{cpw} = 402$~nH/m determined by its geometry. The $m^{\rm th}$ standing wave mode in the CPW-cable-CPW interconnect can be modeled as a lumped element series $LC$ resonator~\cite{Pozar}, with parameters given by
\begin{eqnarray}
  L_m &\approx& \frac{1}{2} (\mathscr{L}_{cb} \ell_{cb}+2 \mathscr{L}_{cpw} \ell_{c})={\rm29.6\: nH}, \\
  \omega_m &\approx& m \omega_{\rm FSR},\\
  C_m &=& \frac{1}{\omega_m^2L_m}.
\end{eqnarray}
Note that for the cable test experiments in section~\ref{sec:wirebond}, there is only one segment of CPW line, therefore $L_m \approx \frac{1}{2} (\mathscr{L}_{cb} \ell_{cb}+\mathscr{L}_{cpw} \ell_{c})$; and the 1.2~mm diameter Al/NbTi cables have a specific capacitance $\mathscr{C}_{cb} = 94$~pF/m and a specific inductance $\mathscr{L}_{cb} = 235$~nH/m.

Similar to Refs.~\onlinecite{Chen2014,Zhong2019,Zhong2021}, the coupling between $Q_i^n$ and the interconnect via the coupler $G_i^n$ is modeled by a tunable inductance given by~\cite{Chen2014,Geller2015}
\begin{equation}\label{M}
  M^n_i = \frac{L_{g}^2}{2L_{g}+L_w+L_{T,i}^n/\cos\delta_i^n},
\end{equation}
where $\delta_i^n$ is the phase across the Josephson junction of $G_i^n$, $L_{T,i}^n$ is the coupler junction inductance at $\delta_i^n=0$, $L_w \approx 0.06$ nH represents the stray inductance of the CPW line connecting the junction with the two linear inductors $L_{g}$, which cannot be ignored when $L_{T,i}^n$ becomes very close to $2L_{g}$~\cite{Zhong2019}.

In the harmonic limit and assuming weak coupling, the coupling between qubit $Q_i^n$ and the $m^{\rm th}$ standing wave mode is~\cite{Chen2014,Geller2015}
\begin{equation}\label{coupling_m}
  g_{i,m}^n = -\frac{M^n_i}{2} \, \sqrt{\frac{\omega_m \omega_{q,i}^n}{(L_{g}+L_{q,i}^n)(L_{g}+L_m)}},
\end{equation}
where $L_{q,i}^n$ is the qubit $Q_i^n$'s junction inductance and $\omega_{q,i}^n/2\pi$ is $Q_i^n$'s operating frequency.

\subsection{Inter-module quantum state transfer and entanglement generation}\label{sec:inter}
In Fig.~3 in the main text, we have shown the inter-module quantum state transfer and entanglement generation data for the $A$-$B$ interconnect. We have also performed similar measurements on other interconnects, as summarized in table~\ref{interconnects}.
\begin{table}[H]
\begin{center}
\begin{tabular}{|l |c | c| c |c |c |c |c |c |c |c}
  \hline
  \hline
                        & $A$-$B$         &$A$-$C$          &$A$-$D$        &$A$-$E$  & average\\
  \hline
  $\mathcal{F}_{QST}$  &$99.1\pm 0.5\%$  &$97.5\pm 0.7\%$ & $97.9\pm 1.2\%$ & $98.1\pm 0.8\%$ & 98.2\%\\
  $\mathcal{F}_{B}$     &$98.9\pm 0.6\%$ &$98.3\pm 0.5\%$ & $97.9\pm 0.7\%$ & $99.1\pm 0.8\%$ &98.6\%\\
  \hline
  \hline
\end{tabular}
\end{center}
\caption{\label{interconnects} {\bf Inter-module quantum state transfer and entanglement generation.}}
\end{table}

\section{Numerical Simulations}\label{sec:simu}
The qubit-cable-qubit system can be modeled with the following rotating-frame Hamiltonian:
\begin{eqnarray}\label{H}
  H/\hbar &=& \sum_{i=1,2} \Delta\omega_{q,i} \sigma_{i}^\dag \sigma_{i} + \sum_{m} \Delta \omega_{m} a_m^\dag a_m \\
  &&+\sum_{m} g_{1,m} \left (\sigma_{1} a_m^\dag + {\sigma_{1}}^\dag a_m \right ) +\sum_{m} (-1)^m g_{2,m} \left (\sigma_{2} a_m^\dag + {\sigma_{2}}^\dag a_m \right)\nonumber,
\end{eqnarray}
where $\sigma_i$ and $a_m$ are the annihilation operators for the two qubits and the $m^{\rm th}$ standing wave mode respectively, $\Delta\omega_{q,i}$ and $\Delta\omega_m$ are the qubit and standing wave mode frequency detuning with respect to the rotating frame frequency, respectively. For simplicity, we can set the rotating frame frequency at the communication mode frequency. Note the sign of $g_{2,m}$ alternates with the mode number $m$ due to the parity dependence of the standing wave mode~\cite{Pellizzari1997,Vogell2017}.

The far detuned standing wave modes can be safely omitted in the simulation as long as $| \Delta \omega_{m}| \gg |g_{i,m}|$. Because the free spectral range of 440~MHz is two orders of magnitude larger than $g_{i,m}\sim 5$~MHz in this experiment, we only keep the communication mode $R$ and omit all the other modes in the simulation, simplifying the Hamiltonian to
\begin{eqnarray}\label{Hsim}
  H/\hbar &=& \sum_{i=1,2} \Delta\omega_{q,i} \sigma_{i}^\dag \sigma_{i}+g_{1} \left (\sigma_{1} a_R^\dag + {\sigma_{1}}^\dag a_R \right ) +  g_{2} \left (\sigma_{2} a_R^\dag + {\sigma_{2}}^\dag a_R \right),
\end{eqnarray}
here $a_R$ is the annihilation operator for the communication mode $R$.
Decoherence is taken into account using the Lindblad master equation. The quantum state evolution is calculated using QuTiP~\cite{Johansson2012}.

\section{Quantum state and process tomography}\label{sec:tomo}
Quantum state tomography~\cite{Steffen2006} can fully characterize a quantum state by applying various, typically perpendicular operations to the quantum state and measuring the corresponding outcome, from which one can numerically reconstruct the density matrix of the quantum state. Here we apply the gate set $\{I, X/2, Y/2\}$ to each qubit before the simultaneous readout of all qubits; the measured probabilities are corrected for readout errors, then we use CVX~\cite{cvx}, a Matlab package for specifying and solving convex programs, to reconstruct the density matrix while constraining it to be Hermitian, unit trace and positive semidefinite. The single-shot simultaneous readout of the qubits is repeated $3\times10^3$ times (for 12 qubits measurement we repeat $6\times10^3$ times) to obtain the measured probabilities; the state tomography is run repeatedly, in each repeat we reconstruct the density matrix and obtain the state fidelity. The fidelities and uncertainties of the quantum states correspond to the mean and standard deviation of 50 repeated measurements.

Quantum process tomography~\cite{Neeley2008} for the state transfer is carried out by preparing input states $\{|0\rangle,(|0\rangle-i|1\rangle)/\sqrt{2}, (|0\rangle+|1\rangle)/\sqrt{2}, |1\rangle\}$ in the transmitting qubit, then performing the quantum state transfer process. The corresponding outcome density matrix in the receiving qubit is measured using quantum state tomography as described above. The process matrix is reconstructed using the input and outcome density matrices, using the CVX package to constrain it to be Hermitian, unit trace, and positive semidefinite.

\clearpage

\bibliographystyle{naturemag}
\bibliography{bibliography}